\documentclass[letterpaper,twocolumn,10pt]{article}
\usepackage{usenix}

\usepackage{amsmath}
\usepackage{amssymb}
\usepackage{amsthm}
\newtheorem{proposition}{Proposition}
\usepackage{booktabs}
\usepackage{multirow}
\usepackage{graphicx}
\usepackage{tikz}
\usetikzlibrary{arrows.meta,positioning,calc,patterns}
\usepackage{pgfplots}
\pgfplotsset{compat=1.16}
\pgfplotsset{every axis/.append style={
  tick label style={font=\normalsize},
  label style={font=\normalsize},
  legend style={font=\normalsize},
  no markers,
  line width=0.5pt, tick style={line width=0.4pt}}}
\pgfplotsset{every node near coord/.append style={text=black}}
\usepackage{xspace}
\usepackage[ruled,vlined,linesnumbered]{algorithm2e}
\usepackage{balance}

\newcommand{\tabfont}{\small}

\newcommand{\papertitle}{\textsc{Carol}: Context-Aware Online Learning for Fuzzer Scheduling}

\newcommand{\sys}{\textsc{Carol}\xspace}           
\newcommand{\rr}{phase-aware scheduler\xspace}     
\newcommand{\rrshort}{phase-aware\xspace}
\newcommand{\ctx}{context-aware scheduler\xspace}  
\newcommand{\ctxshort}{context-aware\xspace}
\newcommand{\base}{BandFuzz\xspace}                
\newcommand{\autofz}{\textsc{autofz}\xspace}       
\newcommand{\legion}{\textsc{Legion}\xspace}       

\newcommand{\aflpp}{AFL++\xspace}
\newcommand{\darwin}{DARWIN\xspace}
\newcommand{\honggfuzz}{honggfuzz\xspace}
\newcommand{\mopt}{MOPT\xspace}
\newcommand{\lafintel}{laf-intel\xspace}
\newcommand{\radamsa}{Radamsa\xspace}

\newcommand{\nEngines}{six\xspace}
\newcommand{\nTargets}{nine\xspace}
\newcommand{\nTrials}{ten\xspace}
\newcommand{\campaignHours}{24\xspace}

\newcommand{\nContextFeatures}{15\xspace}
\newcommand{\nRealDefects}{120\xspace}
\newcommand{\nRealPrograms}{five\xspace}
\newcommand{\nRealPublic}{73\xspace}      
\newcommand{\nRealPrivate}{47\xspace}     

\newcommand{\armset}{\mathcal{A}}                  
\newcommand{\narms}{K}                             
\newcommand{\arm}{a}                               
\newcommand{\round}{t}                             
\newcommand{\ctxvec}{\mathbf{x}}                   
\newcommand{\featmap}{\phi}                        
\newcommand{\parm}{\vartheta}                      
\newcommand{\reward}{r}                            
\newcommand{\budget}{T_{I}}                        
\newcommand{\window}{w}                            
\newcommand{\phasethr}{\tau}                       
\newcommand{\explore}{\alpha}                      
\newcommand{\ridge}{\lambda}                       
\newcommand{\rff}{D}                               

\newcommand{\myparagraph}[1]{\vspace{2pt}\noindent\textbf{#1}}

\newcommand{\bestcell}[1]{\textbf{#1}}

\newcommand{\rotninety}[1]{\rotatebox[origin=l]{90}{\footnotesize #1}}

\usepackage{etoolbox}
\usepackage{microtype}
\apptocmd{\thebibliography}{\raggedright}{}{}
\appto\UrlBreaks{\do\a\do\b\do\c\do\d\do\e\do\f\do\g\do\h\do\i\do\j\do\k\do\l\do\m\do\n\do\o\do\p\do\q\do\r\do\s\do\t\do\u\do\v\do\w\do\x\do\y\do\z\do\-}

\definecolor{cBase}{HTML}{999999}   
\definecolor{cM1}{HTML}{E69F00}     
\definecolor{cM2}{HTML}{0072B2}     
\definecolor{cAcc}{HTML}{009E73}    
\definecolor{cWarn}{HTML}{D55E00}   

\begin{document}

\date{}

\title{\Large \bf \papertitle}


\author{
Zirui Liu$^{1}$ \quad
Mengfan Xu$^{1}$ \quad
Juan Zhai$^{1}$ \quad
Shenglong Yao$^{2}$ \quad
Shiqing Ma$^{1}$\\[3pt]
$^{1}$University of Massachusetts Amherst\\
\texttt{\{zliu,mengfanxu,juanzhai,shiqingma\}@umass.edu}\\[2pt]
$^{2}$Georgia Institute of Technology\\
\texttt{syao84@gatech.edu}
}

\maketitle

\begin{abstract}
Ensemble fuzzing runs several fuzzers on one target and relies on a scheduler to divide CPU time among them. Existing ensemble schedulers make these decisions from compact summaries of past performance and decision rules fixed before the campaign. Our measurements point to two limitations. First, summaries of past rewards do not reliably capture how fuzzer performance evolves: after accounting for estimation noise, the agreement between consecutive-window rankings is statistically indistinguishable from the ranking's within-window self-agreement. Second, the signals that predict reward vary across targets: a weighting learned from eight other targets predicts reward worse than one learned on the current target on eight of \nTargets targets.

We then introduce \sys, which uses each fuzzer's current context to guide online scheduling. The context is already available to the dispatch loop and captures how the fuzzer's reward is evolving, how long it has waited or remained on a plateau, what code it is reaching, and the uncertainty in its estimate. We develop two ways to use this context. A domain-guided method identifies whether a fuzzer is \emph{rising} or \emph{rotting} and applies a learning rule tailored to that phase. A learned method models how reward depends on \nContextFeatures\ context signals and uses its predictions and uncertainty to select fuzzers online. Across \nTargets Magma targets, \sys triggers more unique bugs than each of three ensemble-scheduling baselines on every target where they differ and fewer on none, reaching a mean of $34.20$ unique bugs compared with $28.40$, $28.80$, and $29.00$. It gains $11.8\%$ over the strongest baseline on each target and exceeds an oracle that selects the best single fuzzer per target after the fact ($32.00$). An ablation without context eliminates the gain, and the additional bugs are concentrated among those the baselines trigger rarely or never. Run unchanged on \nRealPrograms widely used C++ programs, \sys finds \nRealDefects previously unknown crashing defects, deduplicated by site, fault, and entry point, all reported to their maintainers through each project's stated disclosure channel.
\end{abstract}

\section{Introduction}
\label{sec:intro}

Greybox fuzzing runs a program on mutated inputs, keeps the ones that reach new code and
reports the ones that crash. No single fuzzer wins everywhere: a decade of measurement
has established that a fuzzer's advantage is
target-specific~\cite{klees2018evaluating,hazimeh2020magma,metzman2021fuzzbench}. The
response has been \emph{ensemble fuzzing}: run several fuzzers on one target, share a
seed corpus, and let a scheduler divide CPU time among
them~\cite{chen2019enfuzz,guler2020cupid,osterlund2021collabfuzz,fu2023autofz,shi2025bandfuzz}.
Choosing a fuzzer becomes choosing how to divide the budget, at every free core and for
the whole campaign, and giving time to the wrong fuzzer leaves the ensemble worse than
its best member. It is online resource allocation under uncertainty about how
each fuzzer will perform on this target.

Online learning is particularly well suited to fuzzer scheduling: it makes decisions in microseconds compared with fuzzing turns that last seconds, learns continuously from the current campaign without pretraining on the target, and produces quantities that can be inspected and interpreted. Prior work has already shown the value of this approach in fuzzing. BandFuzz~\cite{shi2024bandfuzz} and LEGION~\cite{legion2026} use online learning for fuzzer scheduling, while \textsc{Fox}~\cite{fox2024} applies online stochastic control within a single fuzzer. Together, these methods establish online learning as a lightweight and effective framework for adaptive fuzzing.

Turning specifically to ensemble scheduling, several prominent schedulers already use online learning. 
\base~\cite{shi2025bandfuzz,shi2024bandfuzz} uses Thompson sampling to allocate effort among fuzzers, \textsc{Legion}~\cite{legion2026} uses a modified upper-confidence-bound rule over multiple performance signals, and \textsc{autofz}~\cite{fu2023autofz} selects fuzzers based on a preparation window. Earlier systems established the ensemble and seed-sharing setting~\cite{guler2020cupid,chen2019enfuzz}. However, to the best of our knowledge, existing schedulers share two limitations. First, they ultimately rank each fuzzer using a compact summary of its past rewards, which omits information about its current situation and how its performance is changing. When we divide a campaign into twelfths, the rankings in two adjacent time windows have a mean Kendall correlation of only $0.059$, where a higher correlation indicates greater agreement in the fuzzer ordering. For comparison, two random halves of the same window have a correlation of $0.057$. Thus, the ranking from one window predicts the next no better than two noisy estimates formed within the same window (\S\ref{sec:motivation:decay}). Second, existing schedulers use a decision rule fixed before the campaign and shared across targets. Our measurements show that the importance of individual signals varies substantially across targets (\S\ref{sec:ctx:features}). These observations motivate learning how each fuzzer's current context predicts its next reward on the target being fuzzed.

\textbf{Our key insight.} A fuzzer's reward history tells only part of the story. What it will find next also depends on its current context: it may be stuck on a plateau, have just received new seeds from its peers, or be exploring code it has not seen before. The dispatch loop already observes signals that distinguish these situations. For example, in the $90$ minutes before a rarely triggered bug, plateau and freshness signals change sharply even while the coverage rate remains below its campaign average. This behavior is widespread: a fuzzer's reward typically rises as it explores new ground and then decays toward a plateau, with the raw per-turn rate falling in $96.7\%$ of $1{,}841$ (campaign, fuzzer) pairs across all \nTargets targets (\S\ref{sec:motivation:decay}).

This motivates developing online learning methods that use each fuzzer's current context to predict its next reward and guide better scheduling decisions. We explore two ways to incorporate this context into online learning. \textbf{Method~I} (\S\ref{sec:rr}) encodes domain knowledge about how fuzzing rewards evolve: it classifies each fuzzer as \emph{rising} or \emph{rotting} from its recent rewards and applies a learning rule tailored to that phase~\cite{metelli2022rising,seznec2020single}. \textbf{Method~II} (\S\ref{sec:ctx}) automatically learns how reward depends on \nContextFeatures\ signals describing different aspects of each fuzzer's current situation, and uses the resulting predictions and uncertainty to select fuzzers online. Comparing the two reveals how far domain-guided context can go and what is gained by learning the context-reward relationship directly. \sys uses Method~II as its main scheduler, with Method~I as the domain-guided comparison. 

To summarize, \myparagraph{the main contributions are as follows:}

\begin{itemize}

\itemsep0pt

\item \textbf{Empirical evidence that fuzzer scheduling requires context.} We show that a compact summary of past rewards does not fully capture a fuzzer's evolving situation. The raw per-turn reward falls on $96.7\%$ of $1{,}841$ (campaign,~fuzzer) pairs across all \nTargets targets. At the decision timescale, rankings in adjacent reward windows have nearly the same agreement as two estimates formed within the same window, showing that past rewards alone provide little information about how the ranking will evolve (\S\ref{sec:motivation:decay}).

\item \textbf{Two context-aware online-learning methods for fuzzer scheduling.} Method~I incorporates domain knowledge by identifying whether each fuzzer is \emph{rising} or \emph{rotting} and applying a learning rule tailored to that phase. Method~II automatically learns how reward depends on \nContextFeatures\ signals describing different aspects of each fuzzer's current situation, using the resulting predictions and uncertainty to select fuzzers online. A model fitted on eight targets predicts the ninth worse than one learned within that target on eight of \nTargets targets, and the two rank a different fuzzer first on a median $51.0\%$ of decisions (\S\ref{sec:ctx:features}). A construction also shows that context can be necessary: policies using only each fuzzer's own pull counts and realized rewards can incur linear regret when an observable context signal changes its future reward (\S\ref{sec:ctx:assumptions}).

\item \textbf{Ground-truth evaluation of context-aware scheduling.} Across \nTrials \campaignHours-hour campaigns per configuration on \nTargets Magma targets, \sys triggers more unique bugs than each of three ensemble-scheduling baselines on every target where they differ and fewer on none, with summed gains of $20.4\%$, $18.8\%$, and $17.9\%$. An oracle selecting the best of the \nEngines fuzzers per target reaches $32.00$ unique bugs, compared with $34.20$ for \sys. An ablation without context eliminates the gain (\S\ref{sec:eval}, \S\ref{sec:ablation}).

\item \textbf{\nRealDefects previously unknown crashing defects in \nRealPrograms C++ programs.} Running the same \sys scheduler and configuration on MNN, Arm NN, Krita, ncnn, and FreeCAD finds \nRealDefects previously unknown crashing defects, all reported to their maintainers through each project's stated security channel (\S\ref{sec:eval:realworld}).

\end{itemize}

\section{Background}
\label{sec:background}

\subsection{Ensemble fuzzing}
\label{sec:background:ensemble}

\myparagraph{From fixed ensembles to adaptive allocation.}
\textsc{EnFuzz}~\cite{chen2019enfuzz} demonstrated that a diverse set of
seed-sharing fuzzers can outperform its individual members.
\textsc{Cupid}~\cite{guler2020cupid} searches offline for an effective ensemble
composition, while \textsc{CollabFuzz}~\cite{osterlund2021collabfuzz} provides a
broker that coordinates fuzzers under pluggable policies. These systems established
the value of combining complementary fuzzers and sharing their discoveries.

Later systems adapt the allocation during the campaign.
\textsc{autofz}~\cite{fu2023autofz} alternates between a preparation phase that
measures current performance and a focus phase that favors the leaders.
\base~\cite{shi2025bandfuzz,shi2024bandfuzz} maintains a Beta posterior over each
fuzzer's coverage reward and uses Thompson sampling~\cite{thompson1933likelihood}
to select the next fuzzer.
\textsc{RCFuzzer}~\cite{mo2025rcfuzzer} samples a posterior over a
branch-difficulty-weighted coverage score, while
\textsc{AutoFuzz}~\cite{gao2025autofuzz} schedules fuzzer-and-sanitizer pairs with
an estimator for rewards that may change over time.
\textsc{Legion}~\cite{legion2026} combines several performance dimensions using a
modified upper-confidence-bound rule and updates their weights each round.
Together, these systems show that online scheduling can adapt an ensemble using
coverage, recent performance, and complementary measures of fuzzing progress.

\myparagraph{Contextual learning within a fuzzer.}
Context already guides decisions at a finer granularity.
Methods have used online learning to select seeds, mutation operators, and
energy within one fuzzer~\cite{woo2013scheduling,yue2020ecofuzz,lyu2019mopt,
karamcheti2018adaptive,bohme2016aflfast,wang2021aflhier,
karamcheti2018contextual}.
\textsc{CMFuzz}~\cite{wang2021cmfuzz} selects mutation operators from properties of
the current seed, \textsc{Cerebro}~\cite{li2019cerebro} prioritizes seeds using the
code they reach, and \textsc{ReFuzz}~\cite{refuzz2026} models the provenance of
tests reused across processors.
\textsc{Meuzz}~\cite{chen2020meuzz} learns from program and seed features when
choosing work for a concolic engine. In each case, the context describes the input,
seed, or operation being scheduled.

These schedulers differ in how they construct their scores, but each ultimately
ranks a fuzzer from its own observed output. Ensemble scheduling presents a
different context: while a fuzzer waits, its peers change its future opportunity
by adding seeds to the shared store. CAROL represents this live situation using
signals already available to the dispatch loop and learns how they predict the
fuzzer's next reward on the current target. To our knowledge, CAROL is the first
ensemble scheduler to learn this per-fuzzer context-to-reward relationship online.
A construction in \S\ref{sec:ctx:assumptions} formalizes why a fuzzer's own pull
counts and realized rewards can be insufficient under this shared-store coupling.

\myparagraph{Learning with changing rewards.}
Online-learning methods provide several models for rewards that move over time.
Sliding-window and discounted estimators track changes of unrestricted
direction~\cite{garivier2011swucb,russac2019weighted}, while
rising~\cite{metelli2022rising} and
rotting~\cite{levine2017rotting,seznec2020single} models capture
one-directional change. Contextual regression instead predicts reward from observed
features~\cite{agarwal2014taming,foster2020beyond,bietti2021bakeoff}.
These lines provide the two foundations used in this paper: Method~I supplies the
reward phase from domain expertise, while Method~II learns the complete
context-to-reward relationship from the campaign. The next subsection examines how
fuzzing rewards move and why phase is useful but not sufficient.

\subsection{Reward dynamics as context}
\label{sec:motivation:decay}

\begin{figure}[t]
\centering
\pgfplotsset{
  curve/.style={
    width=0.46\columnwidth, height=2.9cm,
    xmin=0.6, xmax=10.4, xtick={1,5,10}, xticklabels={start,mid,end},
    ymin=0, ytick=\empty,
    xlabel={campaign}, xlabel style={font=\normalsize, yshift=2pt},
    axis lines*=left, tick align=outside,
    ymajorgrids, grid style={draw=black!12},
    xticklabel style={font=\normalsize}, title style={font=\normalsize},
  }}
\begin{tikzpicture}
\begin{axis}[curve, ylabel={mean reward}, ylabel style={font=\normalsize},
             title={poppler, \texttt{mopt}}, ymax=0.45]
\addplot[cM2, line width=1.1pt] coordinates
  {(1,0.182) (2,0.315) (3,0.251) (4,0.346) (5,0.404) (6,0.273) (7,0.287) (8,0.266)
   (9,0.185) (10,0.212)};
\end{axis}
\end{tikzpicture}\hfill
\begin{tikzpicture}
\begin{axis}[curve, title={libxml2, \texttt{darwin}}, ymax=0.16]
\addplot[cWarn, dashed, line width=1.1pt] coordinates
  {(1,0.144) (2,0.138) (3,0.139) (4,0.041) (5,0.090) (6,0.043) (7,0.034) (8,0.021)
   (9,0.023) (10,0.004)};
\end{axis}
\end{tikzpicture}
\\[2pt]
\begin{tikzpicture}
\begin{axis}[
  width=0.93\columnwidth, height=3.1cm,
  ybar, bar width=5pt,
  ymin=0, ymax=28,
  ylabel={\% of raw rate retained},
  ylabel style={font=\normalsize, yshift=-6pt},
  symbolic x coords={poppler,libxml2,libsndfile,openssl,lua,sqlite3,libtiff,libpng,php},
  xtick=data,
  xticklabel style={font=\normalsize, rotate=30, anchor=north east},
  yticklabel style={font=\normalsize},
  ytick=\empty,
  nodes near coords, every node near coord/.append style={font=\normalsize, /utils/exec=\color{black}, yshift=-1pt},
  nodes near coords style={/pgf/number format/precision=1, /pgf/number format/fixed},
  enlarge x limits=0.07,
  axis lines*=left,
  ymajorgrids, grid style={draw=black!12},
  tick align=outside,
]
\addplot[ybar, draw=black!60, fill=cM2!45] coordinates
  {(poppler,23.8) (libxml2,4.5) (libsndfile,4.0) (openssl,3.6) (lua,3.1) (sqlite3,3.0)
   (libtiff,0.7) (libpng,0.0) (php,0.0)};
\end{axis}
\end{tikzpicture}
\caption{\emph{Left}, the shape \S\ref{sec:motivation:decay} describes: \texttt{mopt} on
poppler rises as it covers code its peers reached, then rots. \emph{Right}, a case it
does not: \texttt{darwin} on libxml2 never rises. Each curve is one fuzzer's reward,
averaged over that target's campaigns and over each tenth of its own turns; Method~I
assumes the left panel. \emph{Below}, the fall across the suite: the median
(campaign,~fuzzer) pair's raw per-turn rate over a campaign's second half, as a percentage
of its first, latched before any scaling.}
\label{fig:curve}
\label{fig:decay}
\end{figure}
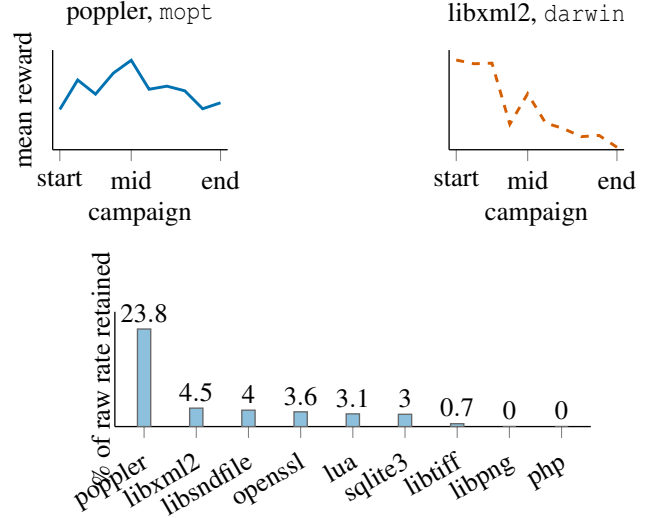

\myparagraph{Seed sharing creates changing opportunities.}
While a fuzzer waits, its peers continue adding seeds to the shared store. When the
fuzzer next runs, those seeds may expose code it has not explored itself, allowing its
reward to rise as it converts the new opportunities into coverage. Its reward then
falls as those opportunities are exhausted and the shared corpus approaches a plateau.
This interaction differs from a recovering-reward model, where recovery depends only
on idle time~\cite{pikeburke2019recovering}: under seed sharing, it also depends on
which peers ran and what they found.

The resulting trajectory is useful but not universal.
Figure~\ref{fig:curve} shows both cases. On poppler, \texttt{mopt} first rises and
then rots as the newly available coverage is consumed. On libxml2,
\texttt{darwin} begins by declining and never enters a rising phase.
We call these phases \emph{rising} and \emph{rotting}, following online-learning
models of one-directional reward change~\cite{metelli2022rising,
levine2017rotting,seznec2020single}. Method~I uses this phase as context supplied
by domain expertise. The contrasting curve shows why phase alone cannot describe
every fuzzer and target.

\myparagraph{Reward decay across the benchmark.}
The decline extends beyond the two examples. The raw per-turn rate falls in
$96.7\%$ of $1{,}841$ (campaign,~fuzzer) pairs across all \nTargets targets, ranging
from $88.6\%$ to $100\%$ by target. The median pair retains $23.8\%$ of its
first-half rate on poppler and less than $5\%$ on each of the other eight targets,
including no retained rate on libpng and php (Figure~\ref{fig:decay}).

We compute this rate from the raw weighted contribution underlying the reward,
before normalization, and separately for each (campaign,~fuzzer) pair. The
measurement therefore describes the signal consumed by the schedulers rather than
a change introduced by its running scale. Coverage rewards are exhaustible by
construction; the measurement establishes how consistently and how sharply that
exhaustion appears for individual fuzzers inside a seed-sharing ensemble.

\myparagraph{Rankings from reward histories.}
Schedulers act on the ordering of fuzzers rather than on a decline shared by all of
them. We therefore divide each campaign into equal wall-clock windows and rank the
fuzzers by the raw rate of the turns they complete in each window. At the
half-campaign width, consecutive-window rankings have a mean Kendall correlation of
$0.081$, while two estimates formed from random halves of the same window agree at
$0.085$. Their paired difference is $0.004$ with a 95\% interval of
$[-0.08, 0.09]$. The same pattern appears at quarter- and twelfth-campaign widths
(\S\ref{sec:appendix:attenuation}).

After accounting for estimation noise, the agreement between consecutive-window
rankings is statistically indistinguishable from the ranking's within-window
self-agreement. Summaries of past reward therefore do not reliably capture how the
fuzzer ordering evolves across the measured windows.

\myparagraph{From phase to full context.}
Reward direction supplies one useful context: it distinguishes a fuzzer whose
reward is rising from one whose reward is rotting toward a plateau. Method~I encodes
that distinction using domain expertise. Direction alone does not describe how long
the fuzzer has waited, how long it has remained on a plateau, what code it is
reaching, where the campaign stands, or how well its reward estimate is determined.
Method~II keeps these signals and learns their relationship to the next reward
online. The two methods therefore answer the same scheduling question with
different amounts of context: one is told which context matters, while the other
learns it from the campaign.

\section{Problem}
\label{sec:background:problem}

\myparagraph{Setting.}
We consider a fixed wall-clock fuzzing campaign on one target with a pool of
$\narms$ fuzzers, $\armset = \{1,\dots,\narms\}$. Each fuzzer is one arm.
All fuzzers instrument the same target build and exchange interesting inputs through
a shared seed store. Whenever a core becomes available, the scheduler selects one
fuzzer to run on it.

\myparagraph{Available observations.}
The scheduler makes each decision from information produced during the campaign and
already maintained by the dispatch loop: coverage of the shared map, turn timing,
observed crashes, static properties of the code reached, and uncertainty estimated
from prior observations. Ground-truth bug identities are reserved for evaluating
completed campaigns.

\myparagraph{Threat model.} We assume an attacker who supplies one untrusted file to a
parser, such as a model, an image, a mesh or a document. The defects we seek are the
memory-safety and input-validation failures on that path: out-of-bounds reads and writes,
dereference of a pointer the file leaves null, use of uninitialized or freed memory, and
an unvalidated quantity from the file used as an index, a divisor or an allocation
size.

\myparagraph{Objective.}
We evaluate a completed campaign by the number of distinct known bugs it triggers.
This metric measures the security outcome directly and counts repeated triggers of
the same defect once, following established guidance for fuzzing
evaluations~\cite{klees2018evaluating}. During the campaign, coverage supplies the
dense feedback needed for online scheduling. After the campaign, Magma's per-bug
oracle identifies the known defects that were triggered. We also report edge
coverage, but use distinct bugs for comparisons because coverage-based and bug-based
evaluations can disagree on which fuzzer is better~\cite{bohme2022reliability}.
Thus, coverage is the online learning signal and unique bugs are the evaluation
objective.

\myparagraph{Design goals.}
\begin{itemize}
\itemsep0pt
\item[\textbf{G1}] \emph{Encode general structure; learn target-specific behavior.}
  The design should make this boundary explicit: general structure in reward
  evolution may be encoded, while target-specific relationships between context
  and reward are learned within the campaign.
\item[\textbf{G2}] \emph{Learn the allocation.}
  CPU shares should follow from online estimates of each fuzzer's next reward and
  the evidence behind them, without fuzzer-specific shares fixed before the
  campaign.
\item[\textbf{G3}] \emph{Deploy on a new target.}
  Every decision should use signals available to the dispatch loop and information
  learned during the current campaign, with no bug oracle, source annotations,
  benchmark harness, or target-specific pretraining.
\item[\textbf{G4}] \emph{Tie design choices to measured outcomes.}
  Questions about predictive information should be evaluated on held-out reward,
  while choices that affect the security outcome should be evaluated on
  full-campaign unique bugs.
\end{itemize}

\section{Method I: A Single Context Guided by Domain Expertise}
\label{sec:rr}

Method~I represents each fuzzer's context by the phase of its reward trajectory.
Domain expertise defines two scheduling phases: \emph{rising} for a new or
upward-trending reward sequence, and \emph{rotting} for a flat or
downward-trending sequence. The phase enters online learning by selecting the
estimator used to predict the fuzzer's next reward. Within that phase, the
scheduler learns the reward level and its uncertainty from the current campaign.
Method~I therefore uses a single context variable---reward phase---whose meaning
is supplied by domain expertise.

This section builds the scheduler in three steps.
\S\ref{sec:rr:dispatch} defines how a scheduling decision becomes a fuzzing turn
and a reward observation. \S\ref{sec:rr:reward} defines that reward and explains
why its changing mean carries scheduling information.
\S\ref{sec:rr:phase} converts the recent reward sequence into a phase, applies
the corresponding estimator, and selects the next fuzzer.

\subsection{The dispatch loop}
\label{sec:rr:dispatch}

\begin{figure}[t]
\centering
\begin{tikzpicture}[
  x=1mm, y=1mm,
  box/.style={draw, rounded corners=1pt, align=center, inner sep=1.5pt,
              font=\normalsize},
  eng/.style={box, minimum width=23mm, minimum height=5mm, fill=cM1!12},
  wide/.style={box, minimum width=76mm, minimum height=6mm},
  inner/.style={box, minimum width=36mm, minimum height=8mm, fill=white},
  lab/.style={font=\normalsize, align=center, inner sep=1pt},
  arr/.style={-{Stealth[length=1.4mm]}}
]

\node[wide, fill=black!3, anchor=north] (workers) at (0,0)
  {workers, one per core};

\node[box, minimum width=76mm, minimum height=17mm, fill=cM2!14,
      anchor=north] (sched) at (0,-9) {};
\node[anchor=north, font=\normalsize] at ([yshift=-1mm]sched.north)
  {\textbf{scheduler}};
\node[box, minimum width=72mm, minimum height=8.5mm, fill=white,
      anchor=north] at ([yshift=-6.8mm]sched.north)
  {\textbf{phase-aware}\\reward phase $\to$ estimator $\to$ fuzzer};

\node[eng, anchor=north west] (e1) at (-37,-33.5) {\aflpp};
\node[eng, anchor=north] (e2) at (0,-33.5) {\darwin};
\node[eng, anchor=north east] (e3) at (37,-33.5) {\honggfuzz};
\node[eng, anchor=north west] (e4) at (-37,-39.5) {\mopt};
\node[eng, anchor=north] (e5) at (0,-39.5) {\lafintel};
\node[eng, anchor=north east] (e6) at (37,-39.5) {\radamsa};

\node[wide, fill=cAcc!8, anchor=north] (store) at (0,-47)
  {shared seed store $\cdot$ coverage map $\cdot$ evaluator};

\draw[arr] (workers.south) -- (sched.north);
\draw[arr] (sched.south) -- (0,-33.5);
\draw[arr] (e5.south) -- (store.north);

\draw[arr] (store.east) -- (43,-50) -- (43,-17.5) -- (sched.east);
\end{tikzpicture}
\caption{Method~I in the dispatch loop. A worker requests a fuzzer from the
phase-aware scheduler, executes its turn, and returns the resulting reward through
the shared evaluator.}
\label{fig:overview}
\end{figure}
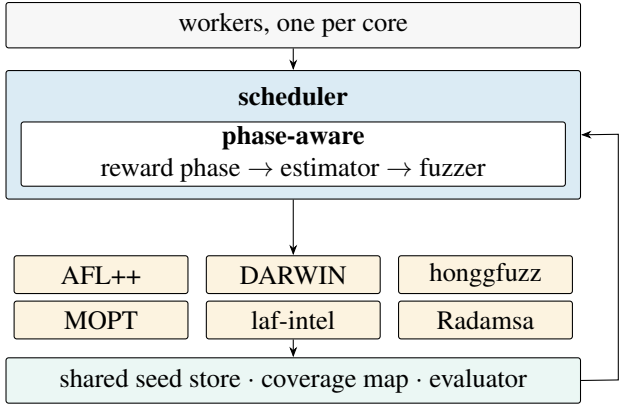

A \emph{worker} is the process that owns one core and executes fuzzing turns.
The \emph{scheduler} receives a request whenever a worker becomes available and
selects the fuzzer for its next turn. Method~I occupies this scheduling component
in Figure~\ref{fig:overview}; the workers, fuzzers, shared seed store, coverage
map, and evaluator form the surrounding execution substrate.

One worker runs on each core. After receiving fuzzer $\arm$, the worker executes
$c_\arm = \lfloor \budget / \bar{d}_\arm \rfloor$ of its internal cycles, where
$\bar{d}_\arm$ is the running mean duration of one such cycle and
$\budget = 120$\,s is the target duration of a turn. It then sends the interesting
inputs to the shared evaluator and reports the resulting reward. Selection
requests are serialized, while the selected turns run concurrently and their
rewards arrive when they finish.

For each request, Method~I classifies every fuzzer's reward phase, computes the
corresponding reward estimate and uncertainty, independently perturbs each
estimate, and selects the largest. Repeating this decision across workers turns
individual selections into a CPU allocation: a fuzzer's expected share is the
probability that its perturbed estimate ranks first. Its share therefore follows
from the phase, reward evidence, and uncertainty accumulated during the campaign,
as required by goal~\textbf{G2}. \S\ref{sec:rr:phase} defines this selection rule,
and \S\ref{sec:eval:rq3} measures the resulting allocation.

\subsection{How reward guides scheduling}
\label{sec:rr:reward}

Reward links a scheduling decision to the progress produced by its fuzzing turn.
If the next reward of every fuzzer were known, the scheduler could select the most
productive next turn directly. Online learning estimates these unknown rewards from
observed turns, so the structure assumed for reward evolution determines how past
evidence informs the next choice.

Every scheduler in our comparison receives the coverage reward defined by
\base~\cite{shi2025bandfuzz,shi2024bandfuzz}. For each new edge covered during
turn $\round$, the evaluator adds $\round-\round_p$, where $\round_p$ is the round
in which the edge's predecessor block was first discovered. The sum is min--max
normalized to $[0,1]$ using running campaign-wide bounds shared by all fuzzers.
Using one reward definition holds the feedback constant while the scheduling rule
changes.

\base summarizes this reward with a Beta posterior whose mean is stationary within
each two-hour interval and resets the posterior at the interval boundary. The
measurements in \S\ref{sec:motivation:decay} show that reward evolves within this
timescale. Method~I represents that evolution directly: the recent reward trend
determines the phase, and the phase determines the estimator for the next reward.
Because most turns contribute no new edge while occasional turns produce large
bursts, the phase test uses reward ranks rather than magnitudes.

\subsection{Phase and estimator}
\label{sec:rr:phase}

\begin{algorithm}[t]
\footnotesize
\SetAlgoLined
\DontPrintSemicolon
\SetKwInOut{Params}{Params}
\Params{trend window $\window=84$, phase threshold $\phasethr=3.5$,
        exploration $\explore=1$; horizons
        $\mathcal{H} = \{2,4,8,\dots\} \cup \{\window\}$}
\KwIn{per-fuzzer reward windows $R_\arm$, pull counts $n_\arm$, decision count $t$}
\KwOut{one fuzzer to run on the calling worker's core}
\BlankLine
\ForEach{fuzzer $\arm \in \armset$}{
  \tcp{a fuzzer without a full window is treated as rising}
  $\mathrm{rising}_\arm \leftarrow |R_\arm| < \window \ \textbf{ or } \
     \textsc{MannKendallZ}(R_\arm) > \phasethr$
    \tcp*{rank trend, tie- and continuity-corrected}
  \eIf{$\mathrm{rising}_\arm$}{
    \tcp{rising: one-step forecast, widened by what the trend leaves open}
    $\hat{s}_\arm \leftarrow \textsc{TheilSenSlope}(R_\arm)$\;
    $v_\arm \leftarrow \mathrm{mean}(R_\arm) + \hat{s}_\arm
       + \explore\sqrt{\log t \cdot \widehat{\mathrm{var}}(\hat{s}_\arm)}$\;
  }{
    \tcp{rotting (a plateau counts as rotting): best sustained level, horizon by maximum}
    $v_\arm \leftarrow \max_{\,h \,\in\, \mathcal{H},\ h \le |R_\arm|}
       \Bigl[\mathrm{mean}\bigl(\text{last } h \text{ of } R_\arm\bigr)
       + \explore\sqrt{\log t / h}\Bigr]$\;
  }
  $s_\arm \leftarrow \sqrt{\widehat{\mathrm{var}}(R_\arm)/|R_\arm| + (1/12)/n_\arm}$\;
  $\epsilon_\arm \sim \mathcal{N}(0,1)$;\quad
  $\tilde{v}_\arm \leftarrow v_\arm + \explore\,\epsilon_\arm s_\arm$
    \tcp*{drawn per worker}
}
\Return $\arg\max_{\arm \in \armset} \tilde{v}_\arm$\;
\caption{The \rr.}
\label{alg:rr}
\end{algorithm}

\myparagraph{Phase classification.}
For fuzzer $\arm$, $R_\arm$ contains its $\window=84$ most recent rewards.
Until the window is full, Method~I initializes the fuzzer's phase as
\emph{rising}. Once the window is full, it computes the standardized
Mann--Kendall statistic with continuity and tie
corrections~\cite{mann1945nonparametric,kendall1948rank}. The fuzzer is
\emph{rising} when this statistic exceeds $\phasethr=3.5$ and
\emph{rotting} otherwise.

Mann--Kendall uses the ordering of reward pairs, so the magnitude of an isolated
coverage burst does not dominate the phase classification. The binary context
groups flat and downward-trending sequences as rotting because both use the same
recent-level estimator in the next step. The phase therefore records the
decision-relevant distinction between an upward trend and a reward that has
stopped rising.

\myparagraph{Phase-specific reward estimates.}
The reward phase is the context input to Method~I: it selects the estimator that
produces fuzzer $\arm$'s index $v_\arm$. For a rising fuzzer, Method~I combines
the recent reward level with a robust estimate of its upward trend:
\[
v_\arm^{\mathrm{rise}}
  = \operatorname{mean}(R_\arm) + \hat{s}_\arm
    + \explore\sqrt{\log t\,
      \widehat{\operatorname{var}}(\hat{s}_\arm)},
\]
where $\hat{s}_\arm$ is the Theil--Sen
slope~\cite{sen1968estimates}. The level and slope capture the direction in which
the reward is moving, while the final term increases the index when that direction
is uncertain, following the rising-reward construction of
Metelli et al.~\cite{metelli2022rising}.

For a rotting fuzzer, Method~I estimates the strongest reward level sustained over
a ladder of recent horizons:
\[
v_\arm^{\mathrm{rot}}
  = \max_{\substack{h \in \mathcal{H}\\h \le |R_\arm|}}
    \left[
      \operatorname{mean}\bigl(\operatorname{last}_h(R_\arm)\bigr)
      + \explore\sqrt{\frac{\log t}{h}}
    \right].
\]
Here $\mathcal{H}=\{2,4,8,\ldots\}\cup\{\window\}$. Each horizon balances a recent reward level against the amount of evidence behind
it. The horizon ladder and its uncertainty term follow
RAW-UCB~\cite{seznec2020single}; Method~I takes the maximum across horizons to
retain the strongest recent evidence, whereas RAW-UCB takes the tightest bound.
\S\ref{sec:appendix:rr} derives how the shipped constants affect the reachable
horizons and the ordering between phases.

\myparagraph{Selection under uncertainty.}
After computing the phase-specific index $v_\arm$, each worker independently
perturbs it:
\[
\tilde{v}_\arm
  = v_\arm + \explore\,\epsilon_\arm s_\arm,
\qquad
\epsilon_\arm \sim \mathcal{N}(0,1),
\]
where
\[
s_\arm
  = \sqrt{
      \frac{\widehat{\operatorname{var}}(R_\arm)}{|R_\arm|}
      + \frac{1/12}{n_\arm}
    }.
\]
The first term estimates uncertainty in the observed reward mean. The second uses
the variance of a uniform prior on $[0,1]$ and decreases with the fuzzer's pull
count $n_\arm$, preserving uncertainty when its observed rewards are sparse or
constant. The worker runs
$\arg\max_{\arm\in\armset}\tilde{v}_\arm$.

Within a phase, the independent perturbations allow uncertainty to change the
fuzzer ordering across concurrent workers, while their scale decreases as the
campaign accumulates evidence. The resulting CPU allocation therefore combines
the domain-guided phase, the phase-specific reward estimate, and the uncertainty
learned online. Method~I uses the same configuration
$\window=84$, $\phasethr=3.5$, and $\explore=1$ on every Magma target.
Algorithm~\ref{alg:rr} summarizes the complete decision rule, and
\S\ref{sec:appendix:rr} derives the behavior of its indices and uncertainty term.

\subsection{Implementation}
\label{sec:impl}

\sys replaces one function of an ensemble dispatch loop, the choice of which fuzzer
runs next. Every scheduler compared in this paper, ours and the three baselines, is
that same loop with a different rule in that one place and everything else held fixed,
\base's own reward definition included.
The pool is \nEngines fuzzers: \aflpp~\cite{fioraldi2020aflpp},
\darwin~\cite{jauernig2023darwin}, \honggfuzz~\cite{honggfuzz},
\mopt~\cite{lyu2019mopt}, \lafintel~\cite{lafintel} and \radamsa~\cite{radamsa}. A
control channel starts, runs and stops each for a bounded number of its own internal
cycles without losing what it has learned, so a turn is a suspension rather than a
restart. \lafintel is an instrumentation pass rather than a mutator: it drives \aflpp\
against a build whose multi-byte comparisons are split, so each target is compiled once
per instrumentation family. All fuzzers publish interesting inputs into one seed store
and import from it before each turn, each input at most once per fuzzer, and every turn's
reward is measured on one neutral build carrying a runtime no fuzzer uses for its own
guidance. Every fuzzed target runs under a fixed address-space limit no healthy target on
these benchmarks approaches, and a persistent-mode fuzzer paired with a leaking harness is
contained at the process that causes it. One configuration ran on every Magma target and
every real-world program; \S\ref{sec:ablation} measures what each setting is worth.

\section{Method II: Context-Aware Online Learning}
\label{sec:ctx}

Method~II models each fuzzer's expected next reward as an unknown function of its
current context:
\[
  \mathbb{E}\!\left[
    \reward_{\round,\arm}\mid\ctxvec_{\round,\arm}
  \right]
  = f_\arm(\ctxvec_{\round,\arm}).
\]
At decision $\round$, every fuzzer $\arm$ presents a context
$\ctxvec_{\round,\arm}$. A model learned from the current campaign predicts its
next reward and the uncertainty of that prediction. The scheduler selects a
fuzzer using these quantities, observes the reward produced by its turn, and
updates that fuzzer's model. Context therefore enters directly as the input to
the learned reward function.

Method~I supplies one context variable---reward phase---from domain expertise.
Method~II instead retains the \nContextFeatures-dimensional context and learns
which combinations of its signals predict reward on the target being fuzzed.

This section builds the method in four steps.
\S\ref{sec:ctx:features} defines the context signals and establishes why they are
retained without a low-rank projection.
\S\ref{sec:ctx:model} learns one context-to-reward function per fuzzer.
\S\ref{sec:ctx:select} combines each prediction with its uncertainty to select
the next fuzzer.
\S\ref{sec:ctx:assumptions} states the concurrent, delayed-feedback setting and
shows why observable context can be necessary when fuzzers interact through the
shared seed store.

\subsection{Context signals}
\label{sec:ctx:features}

\myparagraph{Context definition.}
At each decision, the scheduler constructs one \nContextFeatures-dimensional
context for every fuzzer from quantities already maintained by the dispatch loop.
The signals describe five aspects of its current situation: recent productivity;
reward trend and short-term horizons; plateau duration and freshness; properties
of the code it has recently reached; and the quality of its current estimate.
Table~\ref{tab:features} defines every signal.

Turn-derived signals are latched when the fuzzer's preceding turn ends.
Only the two clock-based signals advance while it waits. The context used to
predict a turn therefore contains no outcome from that turn. All signals are
available on a new target: code properties are read from the instrumented binary,
crashes are ordinary observed crash counts, and no signal uses a bug oracle,
source annotation, or benchmark harness.

\myparagraph{Retaining the context.}
Two rules fixed before the campaigns remove one invariant signal and five exact
arithmetic functions of retained quantities, leaving the \nContextFeatures\ signals
in Table~\ref{tab:features}. In leave-one-campaign-out prediction, the signals that
carry information differ by target, and no tested low-rank projection improves on
the complete context on any target. Method~II therefore retains the full context
(\S\ref{sec:appendix:featsel}, \S\ref{sec:appendix:ablation}).

\myparagraph{Learning the weighting on the current target.}
Method~II learns how the context signals combine from the campaign being
scheduled. We measure the need for this target-specific weighting in two
complementary ways.

First, replaying each decision after replacing one feature by its standardized
mean identifies a different dominant signal across targets
(Table~\ref{tab:decision-impact}).

Second, we compare the same estimator under two sources of training data. One
model is fitted on campaigns from the other eight targets and transferred to the
held-out target; the other is fitted within that target. The transferred
weighting has significantly higher held-out prediction error on eight of the
\nTargets targets after Holm correction and lower error on none. It is worse on
$66.7\%$ to $97.4\%$ of each target's held-out campaigns. The difference reaches
the scheduling decision: the transferred and within-target models select
different top-ranked fuzzers on $26.8\%$ to $60.4\%$
of decisions per target, with a median of $51.0\%$.

These held-out analyses test whether one weighting transfers across targets.
The deployed scheduler initializes its models from the current campaign and
learns the weighting online, satisfying goal~\textbf{G3}.

\myparagraph{Representation boundary.}
Proposition~\ref{prop:blind} uses an ideal indicator of whether the preceding
shared-store publication opened code new to a fuzzer. The implemented context
instead records consequences visible to the dispatch loop, including productivity,
plateau duration, freshness, and properties of the code reached. The proposition
establishes the information need; \S\ref{sec:ablation:context} measures the value
of the context Method~II implements.

\subsection{The model}
\label{sec:ctx:model}

\myparagraph{One online model per fuzzer.}
Method~II maintains a separate reward model for every fuzzer. This allows each
model to learn the fuzzer's persistent productivity and its response to context;
the selected arm already supplies fuzzer identity, so identity is not repeated as
a context feature.

For fuzzer $\arm$, each component of
$\ctxvec_{\round,\arm}\in\mathbb{R}^{\nContextFeatures}$ is standardized using
that fuzzer's running mean and variance and clamped to $\pm5$ standard deviations.
The scheduler then applies the fixed random-Fourier
representation~\cite{rahimi2007random}
\[
  \featmap(\ctxvec)
  =
  \left[
    \sqrt{\frac{2}{\rff}}
      \cos(\omega_1^{\!\top}\ctxvec+b_1),
    \ldots,
    \sqrt{\frac{2}{\rff}}
      \cos(\omega_\rff^{\!\top}\ctxvec+b_\rff),
    1
  \right]^{\!\top},
\]
where the final component is the intercept. We use
$\rff=16$, bandwidth $\sigma=4$,
$\omega_j\sim\mathcal{N}(0,\sigma^{-2}I)$, and
$b_j\sim\mathrm{Unif}[0,2\pi]$. The frequencies and phases are drawn once and
then held fixed.

Each fuzzer carries ridge sufficient statistics initialized at
$A_\arm=\ridge I$ and $b_\arm=0$, with $\ridge=10$. When one of its turns
completes with representation $\featmap_\arm$ and reward $\reward$, Method~II
updates
\[
  A_\arm
    \leftarrow A_\arm+\featmap_\arm\featmap_\arm^{\!\top},
  \qquad
  b_\arm
    \leftarrow b_\arm+\reward\featmap_\arm.
\]
Its parameter and predicted reward at the next context are
\[
  \parm_\arm=A_\arm^{-1}b_\arm,
  \qquad
  \mu_\arm(\ctxvec_{\round,\arm})
    =\parm_\arm^{\!\top}\featmap(\ctxvec_{\round,\arm}).
\]
The standardizers and models begin with each campaign and are updated only from
that campaign's completed turns, satisfying goal~\textbf{G3}.

At the data volumes reached during a campaign, this representation has
$6.9\%$--$36.1\%$ lower held-out prediction error than a model linear in the raw
context on both targets where the comparison is measured
(\S\ref{sec:ablation:kernel}). In the live comparison, the two representations
differ by $0.20$ mean unique bugs, which the full-campaign test does not
distinguish (\S\ref{sec:ablation:judging}).

\myparagraph{Using the full campaign history.}
Method~II gives equal weight to every completed turn in the current campaign.
Forgetting factors $\gamma=0.995$ and $\gamma=0.99$ both increase held-out
prediction error on all \nTargets targets after Holm correction and reduce it on
none, so Method~II retains the full campaign history. Elapsed time, idle time, and
freshness let a fixed coefficient vector express systematic change, although
residual drift remains on three targets. \S\ref{sec:eval:rq3} places that drift
alongside the direct comparison between the methods; \S\ref{sec:appendix:gamma}
gives the full weighting analysis.

\subsection{Selection}
\label{sec:ctx:select}

\begin{algorithm}[t]
\SetAlgoLined
\DontPrintSemicolon
\SetKwInOut{Params}{Params}
\Params{ridge $\ridge = 10$, random-Fourier width $\rff = 16$, bandwidth
        $\sigma = 4$}
\KwIn{per-fuzzer ridge accumulators $(A_\arm, b_\arm)$, running standardizers}
\KwOut{one fuzzer to run on the calling worker's core}
\BlankLine
\ForEach{fuzzer $\arm \in \armset$}{
  $\ctxvec_\arm \leftarrow \textsc{Standardize}_\arm(\textsc{Context}(\arm))$
    \tcp*{\nContextFeatures\ features, clamped to $\pm 5$}
  $\featmap_\arm \leftarrow
     \bigl[\sqrt{2/\rff}\cos(\omega_j^{\!\top}\ctxvec_\arm + b_j)\bigr]_{j=1}^{\rff}
     \,\Vert\, 1$ \tcp*{$\omega, b$ frozen at start}
  $\parm_\arm \leftarrow A_\arm^{-1} b_\arm$ \tcp*{one model per fuzzer}
  $\mu_\arm \leftarrow \parm_\arm^{\!\top}\featmap_\arm$;\quad
  $u_\arm \leftarrow \sqrt{\featmap_\arm^{\!\top} A_\arm^{-1} \featmap_\arm}$\;
  $\epsilon_\arm \sim \mathcal{N}(0,1)$;\quad
  $\tilde{\mu}_\arm \leftarrow \mu_\arm + \epsilon_\arm u_\arm$
    \tcp*{drawn per worker}
}
$\arm^{\star} \leftarrow \arg\max_{\arm \in \armset} \tilde{\mu}_\arm$;\quad
record $\featmap_{\arm^{\star}}$ for the update\;
\BlankLine
\tcp{on the turn's completion, with realized reward $\reward$}
$A_{\arm^{\star}} \leftarrow A_{\arm^{\star}} +
   \featmap_{\arm^{\star}}\featmap_{\arm^{\star}}^{\!\top}$;\quad
$b_{\arm^{\star}} \leftarrow b_{\arm^{\star}} + \reward\,\featmap_{\arm^{\star}}$\;
\Return $\arm^{\star}$\;
\caption{The \ctx. $A_\arm$ is initialized to $\ridge I$.}
\label{alg:ctx}
\end{algorithm}

\myparagraph{Combining prediction and uncertainty.}
For each fuzzer, Method~II pairs its predicted reward with the ridge width
\[
  u_\arm
  =
  \sqrt{\featmap_\arm^{\!\top}A_\arm^{-1}\featmap_\arm}.
\]
The width is large in representation directions supported by few observations and
shrinks as similar contexts accumulate. It therefore identifies where the current
campaign provides the model with the least evidence.

At each decision, each worker independently draws
\[
  \tilde{\mu}_\arm
  =
  \mu_\arm(\ctxvec_{\round,\arm})
  +
  \epsilon_\arm u_\arm,
  \qquad
  \epsilon_\arm\sim\mathcal{N}(0,1),
\]
for every fuzzer and selects the largest draw. The predicted mean favors fuzzers
expected to produce high immediate reward, while the perturbation keeps fuzzers
with weakly observed contexts in contention. Independent draws also allow
concurrent workers to make different selections from the same model state.

This mean-plus-Gaussian-width rule follows randomized exploration in contextual
online learning~\cite{agrawal2013thompson}. We fix the perturbation multiplier to
one for every target, so its scale is determined by the ridge width at the current
context rather than by a separate fuzzing-specific exploration coefficient.
The exploration term is consequential: removing it reduces mean unique bugs by
$1.00$ per target, or $9.00$ across the \nTargets targets, and reduces edge
coverage by $16\%$. In that ablated configuration, a passed-over fuzzer regains
the highest prediction zero times across all learned decisions
(\S\ref{sec:ablation:settled}, \S\ref{sec:appendix:select}).

\subsection{The setting}
\label{sec:ctx:assumptions}

\myparagraph{Concurrent, coupled decisions.}
At decision $\round$, an available worker chooses one of the $\narms$ fuzzers.
Every fuzzer $\arm$ presents its current context
$\ctxvec_{\round,\arm}$, and the selected fuzzer runs for $\budget$ seconds.
Up to \nEngines worker turns can be in flight at once. A model is updated only
when its turn completes, so later decisions can be made before earlier selections
have produced feedback, and completed turns can arrive out of selection order.

Each completed turn can publish inputs to the shared seed store and thereby change
the opportunities available to every fuzzer. A fuzzer's context can therefore
evolve while it waits, and its next reward reflects both its own recent behavior
and the consequences of its peers' work. Online-learning formulations with context
model reward conditioned on observed context~\cite{li2010linucb,valko2013kernelucb},
while restless and recovering formulations model rewards that evolve over
time~\cite{whittle1988restless,pikeburke2019recovering}. Ensemble fuzzing combines
these properties with concurrent feedback and shared-store coupling.

Method~II models the context-to-reward relationship as fixed within a campaign
while allowing the contexts themselves to evolve after every turn.
\S\ref{sec:ctx:model} measures this choice and quantifies the residual drift that
the context does not capture.

\myparagraph{Immediate-reward objective.}
Method~II selects the fuzzer with the strongest randomized estimate of its next-turn
reward. A long-horizon policy would additionally model how each selection changes
the shared store and all future contexts; Method~II does not estimate that transition
or value function. \S\ref{sec:appendix:setting} examines this design choice using
retained reward and alternative reward definitions.

Shared-store coupling nevertheless makes context necessary for the immediate
decision: a fuzzer's own pull counts and realized rewards do not reveal the
opportunities its peers created while it waited. Proposition~\ref{prop:blind}
isolates this missing information.

\begin{proposition}
\label{prop:blind}
For every horizon $H$, there is an instance satisfying the monotone coupling of
\S\ref{sec:motivation:decay} in which every policy whose decisions use only the
fuzzers' own pull counts and realized rewards has expected cumulative reward
$3H/4$. A policy that additionally observes which fuzzers can exploit the
preceding shared-store publication attains expected cumulative reward $7H/8$.
The history-only policies therefore incur expected regret $H/8$ relative to this
context-informed policy.
\end{proposition}

The construction makes the preceding shared-store publication independently
useful to each of two fuzzers. Pull counts and realized rewards do not identify
which fuzzer can exploit the current publication, while one observable indicator
does. \S\ref{sec:appendix:prop} gives the full construction, proof, and scope.

Method~II uses the dispatch-loop signals of \S\ref{sec:ctx:features}, not the
construction's direct indicator. Replacing those signals with an intercept while
holding the remaining design fixed lowers summed mean unique bugs from $34.20$ to
$28.00$, below \base's $28.40$; the implemented context therefore carries the
entire measured advantage and more (\S\ref{sec:ablation:context}). The proposition
establishes the information requirement, and the full-campaign comparison measures
the deployed scheduler under concurrent, delayed feedback.

\section{Evaluation}
\label{sec:eval}

\subsection{Setup}
\label{sec:eval:setup}

\myparagraph{Benchmark.}
We use the \nTargets-target Magma release~\cite{hazimeh2020magma} and, for each
target, the harness reaching the most injected bugs. Magma's reached and triggered
oracles score completed campaigns but are hidden from every scheduler. Campaigns
start from Magma's seed corpus without a dictionary.

\myparagraph{Campaign protocol.}
Each cell contains \nTrials independent \campaignHours-hour campaigns on
\nEngines cores under the same limits (\S\ref{sec:impl}). Campaigns have distinct
seeds, containers, and corpus directories, exchange no runtime state, and launch
together by target.
Mean cell wall-clock is $24.00$--$24.09$ hours. The $118$ cells comprise $45$
scheduler--target, $54$ standalone-fuzzer--target, nine context-ablation, and ten
alternative-reward cells: $1{,}180$ campaigns and $169{,}920$ core-hours. Equal
wall-clock and core allocation define the budget because scheduling itself changes
each engine's corpus and execution rate.

\myparagraph{Controlled scheduler comparison.}
All five schedulers share the fuzzers, harness, seeds, seed sharing, coverage map,
reward, container, resources, and duration; only allocation changes. We implement
\base~\cite{shi2025bandfuzz}, \autofz~\cite{fu2023autofz}, and
\legion~\cite{legion2026} in the common loop with their stated inputs and
constants, without target-specific tuning. The appendix maps every rule and the
conversion of round-level shares to \nEngines whole cores
(\S\ref{sec:appendix:comparison}); where paper and released code differ, we follow
the paper. Each fuzzer also runs alone under the same budget. For target-level
summaries, the comparator is the highest observed baseline mean on that
target---a post-campaign composite, not one deployable scheduler.
Figure~\ref{fig:single} also gives the analogous standalone-fuzzer composite.

\myparagraph{Outcomes and logs.}
The primary outcome is mean distinct Magma bugs over the \nTrials campaigns, with
standard deviation; edge coverage is secondary. Logs record each decision,
selected fuzzer, reward, and, where produced, context, prediction, and randomized
draw. Each offline analysis states the campaigns and fields it uses.

\myparagraph{Pairwise inference.}
For each target and pair, we report Vargha--Delaney $\hat{A}_{12}$ and a two-sided
Mann--Whitney U test~\cite{mann1947test,arcuri2014hitchhiker}, Holm-corrected over
the \nTargets-target family~\cite{holm1979simple}. A difference is resolved when
its gap reaches its pair-specific permutation resolution and
$p_{\mathrm{H}}<0.05$; otherwise we report direction. An exact two-sided sign test
evaluates the non-tied target directions (\S\ref{sec:appendix:resolution}).

\subsection{RQ1: Unique bugs}
\label{sec:eval:rq1}

\begin{table*}[t]
\centering
\tabfont
\caption{Mean unique Magma bugs over \nTrials \campaignHours-hour campaigns per cell.
The three right-hand schedulers are the baselines, each reimplemented from its own paper
with its own constants and run in this harness on the same \nEngines fuzzers.
$\Delta$ is what the \ctx makes over whichever of the three is strongest on that target,
and $p_{\mathrm{H}}$ a two-sided Mann--Whitney U test of it after Holm correction over
the family of \nTargets targets. \textbf{Bold} marks the highest mean in a row and
additionally marks Method~I when it exceeds all three baselines. In the gain columns,
bold marks a $\Delta$ at or above what its own pair resolves
(\S\ref{sec:ablation:judging}) and $p_{\mathrm{H}} < 0.05$. Each fuzzer alone is
Table~\ref{tab:single}, standard deviations
Table~\ref{tab:main-sd}, edge coverage Table~\ref{tab:main-edges} and every pair's effect
size Table~\ref{tab:main-full}.}
\label{tab:main}
\setlength{\tabcolsep}{2.2pt}
\begin{tabular}{@{}l rr rrr rr@{}}
\toprule
& \multicolumn{2}{c}{this paper} & \multicolumn{3}{c}{baseline} & \multicolumn{2}{c}{gain} \\
\cmidrule(lr){2-3}\cmidrule(lr){4-6}\cmidrule(lr){7-8}
target & \ctxshort & \rrshort & \base & \autofz & \legion & $\Delta$ & $p_{\mathrm{H}}$ \\
\midrule
libpng & \bestcell{4.00} & \bestcell{3.90} & 3.40 & 3.20 & 3.60 & +0.40 & 0.235 \\
libsndfile & 7.00 & 7.00 & 7.00 & 7.00 & 7.00 & +0.00 & 1.000 \\
libtiff & \bestcell{5.00} & \bestcell{4.30} & 3.70 & 4.00 & 3.60 & \bestcell{+1.00} & \textbf{0.0001} \\
libxml2 & \bestcell{3.20} & \bestcell{2.70} & 2.10 & 2.60 & 2.20 & +0.60 & 0.317 \\
lua & \bestcell{1.60} & 1.30 & 1.10 & 1.20 & 1.40 & +0.20 & 1.000 \\
openssl & \bestcell{2.00} & 1.80 & 1.60 & \bestcell{2.00} & 1.60 & +0.00 & 1.000 \\
php & 3.00 & 3.00 & 3.00 & 3.00 & 3.00 & +0.00 & 1.000 \\
poppler & \bestcell{4.20} & 3.20 & 3.10 & 2.80 & 3.60 & +0.60 & 0.478 \\
sqlite3 & \bestcell{4.20} & 3.30 & 3.40 & 3.00 & 3.00 & \bestcell{+0.80} & \textbf{0.029} \\
\midrule
summed mean & \bestcell{34.20} & \bestcell{30.50} & 28.40 & 28.80 & 29.00 & \multicolumn{2}{r@{}}{+11.8\% over the strongest} \\
\bottomrule
\end{tabular}
\end{table*}

\input{figures/fig_curves}

Table~\ref{tab:main} reports the schedulers, Table~\ref{tab:single} the standalone
fuzzers, and Figure~\ref{fig:curves} the four separating trajectories. All
schedulers trigger the same seven bugs on libsndfile and three on php despite
additional reachable sites (Table~\ref{tab:ceilings}); the other seven targets
determine the directions below.

\myparagraph{Method~II against the baselines.}
The \ctx triggers more unique bugs than each baseline on every target where they
differ and fewer on none. Summed across the \nTargets targets, it triggers
$34.20$ mean unique bugs against \base's $28.40$ ($+20.4\%$),
\autofz's $28.80$ ($+18.8\%$), and \legion's $29.00$ ($+17.9\%$).

The strongest-baseline composite sums to $30.60$: Method~II leads it on six
targets, ties on three, and gains $11.8\%$ (exact sign $p=0.031$). The resolved
gaps are libtiff, $5.00$ versus $4.00$
($\hat{A}_{12}=1.00$, $p_{\mathrm{H}}=0.0001$), and sqlite3, $4.20$ versus
$3.40$ ($p_{\mathrm{H}}=0.029$). Every other non-tied effect favors Method~II.
Removing libtiff still leaves a $9.8\%$ gain; normalized by reachable bugs, the
composite reaches $45.4\%$ and Method~II $49.6\%$.

Against \base, \autofz, and \legion individually, the exact sign probabilities
are $0.0156$, $0.031$, and $0.0156$. Every Method~II campaign on libtiff
triggers more bugs than every campaign of its strongest baseline, and all ten
Method~II campaigns on libpng trigger the same four bugs. Restricting the sum to
the six targets where Method~II and the composite differ gives a $19.4\%$ gain.

\myparagraph{Method~I and the comparison between the two designs.}
Method~I sums to $30.50$, improving on \base, \autofz, and \legion by
$7.4\%$, $5.9\%$, and $5.2\%$. Its strongest resolved result is libpng:
$3.90$ versus \autofz's $3.20$ ($\hat{A}_{12}=0.85$,
$p_{\mathrm{H}}=0.023$), where reward decay is sharpest among separating
targets (Figure~\ref{fig:decay}). Method~II improves on Method~I by $12.1\%$,
leading on seven targets and tying on two. The resolved pairwise effects are
sqlite3 ($\hat{A}_{12}=0.88$, $p_{\mathrm{H}}=0.016$) and libtiff
($\hat{A}_{12}=0.80$, $p_{\mathrm{H}}=0.025$).
Against the strongest-baseline composite, Method~I leads on three targets, ties
on two, and trails on four, summing to $30.50$ against $30.60$. Poppler is the
next Method~II--Method~I comparison after the two resolved targets, with
$\hat{A}_{12}=0.835$ and $p_{\mathrm{H}}=0.069$.

\myparagraph{Time to trigger difficult bugs.}
Kaplan--Meier estimates~\cite{hazimeh2020magma,kaplan1958} censor untriggered
campaigns at \campaignHours hours. Method~II establishes medians absent from the
strongest baseline for TIF008 ($11.6$ hours), XML009 ($13.7$), SQL020 ($15.1$),
and XML001 ($22.1$); no bug shows the reverse (Table~\ref{tab:ttb}). Among the
$30$ bugs with medians under both, the median time difference is zero. An
in-budget median requires at least five of ten campaigns to trigger the bug, so
the timing gain comes from making four difficult bugs repeatable within the
campaign budget rather than shifting the common bugs earlier.

\myparagraph{Edge coverage and bug-site reach.}
Method~II covers more edges than the composite on five targets, ties on php, and
covers fewer on three (Table~\ref{tab:main-edges}). On libxml2 it covers $8{,}954$
versus $9{,}210$ edges yet triggers $3.20$ versus $2.60$ bugs. It nevertheless
reaches more injected bug sites than every baseline on libtiff, libxml2, and
sqlite3, ties on the other six, and loses on none
(Table~\ref{tab:main-reached}). The gain therefore tracks additional bug sites,
not broader edge coverage alone.

\subsection{RQ2: Which bugs}
\label{sec:eval:rq2}

\begin{table*}[t]
\centering
\tabfont
\caption{Per-bug campaign counts for libpng, libtiff, and sqlite3:
the two targets with resolved Method~II gains and the target on which
every Method~II campaign triggers the same four bugs. Rows list every
bug triggered by any scheduler in the table, and each entry is the
number of that cell's \nTrials campaigns that trigger it. Dividing a
column sum by \nTrials gives the corresponding mean in
Table~\ref{tab:main}. \textsc{new} marks a bug triggered by no baseline
campaign.}
\label{tab:perbug}
\setlength{\tabcolsep}{5pt}
\begin{tabular}{@{}llrrrrl@{}}
\toprule
target & bug & \base & \autofz & \legion & \ctxshort (this paper) & \\
\midrule
\multirow{5}{*}{libpng}
        & PNG003 & 10 & 10 & 10 & 10 &  \\
        & PNG006 & 10 & 10 & 10 & 10 &  \\
        & PNG007 & 10 & 10 & 10 & 10 &  \\
        & PNG001 & 4 & 2 & 6 & \textbf{10} &  \\
\cmidrule(l){2-7}
        & \emph{mean} & \emph{3.40} & \emph{3.20} & \emph{3.60} & \emph{4.00} & \\
\midrule
\multirow{7}{*}{libtiff}
        & TIF007 & 10 & 10 & 10 & 10 &  \\
        & TIF012 & 10 & 10 & 10 & 10 &  \\
        & TIF014 & 10 & 10 & 10 & 10 &  \\
        & TIF002 & 4 & 10 & 6 & 9 &  \\
        & TIF008 & 2 & 0 & 0 & \textbf{8} &  \\
        & TIF001 & 1 & 0 & 0 & \textbf{3} &  \\
\cmidrule(l){2-7}
        & \emph{mean} & \emph{3.70} & \emph{4.00} & \emph{3.60} & \emph{5.00} & \\
\midrule
\multirow{8}{*}{sqlite3}
        & SQL002 & 10 & 10 & 10 & 10 &  \\
        & SQL014 & 10 & 10 & 10 & 10 &  \\
        & SQL018 & 10 & 10 & 10 & 10 &  \\
        & SQL020 & 3 & 0 & 0 & \textbf{5} &  \\
        & SQL012 & 1 & 0 & 0 & \textbf{3} &  \\
        & SQL013 & 0 & 0 & 0 & \textbf{3} & \textsc{new} \\
        & SQL015 & 0 & 0 & 0 & \textbf{1} & \textsc{new} \\
\cmidrule(l){2-7}
        & \emph{mean} & \emph{3.40} & \emph{3.00} & \emph{3.00} & \emph{4.20} & \\
\bottomrule
\end{tabular}
\end{table*}

Table~\ref{tab:perbug} decomposes the bug-count differences on libpng,
libtiff, and sqlite3. Nine common bugs contribute equally. PNG001 produces
libpng's gain; TIF008 and TIF001, partly offset by TIF002, produce libtiff's;
and SQL020, SQL012, SQL013, and SQL015 produce sqlite3's.
Specifically, Method~II triggers PNG001 in all ten campaigns against \legion's
six. Relative to the strongest baseline on each target, it adds eight campaign
triggers on TIF008 and three on TIF001, offset by one fewer on TIF002; on sqlite3
it adds two each on SQL020 and SQL012, three on SQL013, and one on SQL015.

\myparagraph{Bugs outside the baseline union.}
Method~II triggers three bugs absent from every baseline campaign: SQL013 in
three campaigns, PDF021 in one, and SQL015 in one (Table~\ref{tab:ttb}). Against
each baseline separately, the absent-bug counts are six for \base, eleven for
\autofz, and nine for \legion. The gain therefore includes new triggered-bug
coverage as well as more reliable triggering of shared bugs.

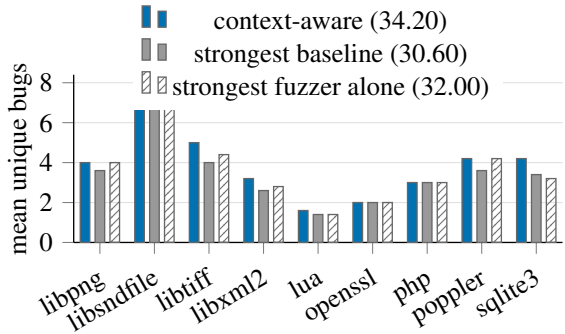
\begin{figure}[t]
\centering
\begin{tikzpicture}
\begin{axis}[
  width=0.95\columnwidth, height=3.8cm,
  ybar, bar width=3.6pt,
  ymin=0, ymax=8.4,
  ylabel={mean unique bugs},
  ylabel style={font=\normalsize, yshift=-4pt},
  symbolic x coords={libpng,libsndfile,libtiff,libxml2,lua,openssl,php,poppler,sqlite3},
  xtick=data,
  xticklabel style={font=\normalsize, rotate=30, anchor=north east},
  yticklabel style={font=\normalsize},
  ytick={0,2,4,6,8},
  enlarge x limits=0.06,
  axis lines*=left,
  ymajorgrids, grid style={draw=black!12},
  tick align=outside,
  legend style={font=\normalsize, at={(0.5,1.45)}, anchor=north, legend columns=1,
                draw=none, /tikz/every even column/.append style={column sep=5pt}},
]
\addplot[ybar, draw=black!60, fill=cM2] coordinates {(libpng,4.00) (libsndfile,7.00) (libtiff,5.00) (libxml2,3.20) (lua,1.60) (openssl,2.00) (php,3.00) (poppler,4.20) (sqlite3,4.20)};
\addplot[ybar, draw=black!60, fill=cBase] coordinates {(libpng,3.60) (libsndfile,7.00) (libtiff,4.00) (libxml2,2.60) (lua,1.40) (openssl,2.00) (php,3.00) (poppler,3.60) (sqlite3,3.40)};
\addplot[ybar, draw=black!60, pattern=north east lines, pattern color=black!55] coordinates {(libpng,4.00) (libsndfile,7.00) (libtiff,4.40) (libxml2,2.80) (lua,1.40) (openssl,2.00) (php,3.00) (poppler,4.20) (sqlite3,3.20)};
\legend{\ctxshort\ (34.20), strongest baseline (30.60), strongest fuzzer alone (32.00)}
\end{axis}
\end{tikzpicture}
\caption{Method~II against two post-campaign composites. For each
target, the baseline bar selects the highest mean among the three
baselines, and the standalone-fuzzer bar selects the highest mean among
the \nEngines fuzzers run alone. The legend reports the sums across
targets; Table~\ref{tab:single} gives every standalone-fuzzer cell.}
\label{fig:single}
\end{figure}

\myparagraph{Against standalone fuzzers.}
Each fuzzer also runs alone under the same protocol
(\S\ref{sec:appendix:single}). The strongest fixed fuzzer sums to $28.40$, and
the post-campaign per-target composite reaches $32.00$ (Figure~\ref{fig:single}).
Method~II reaches $34.20$: $20.4\%$ above the fixed choice and $6.9\%$ above the
composite, leading it on four targets, tying five, and losing none. Sqlite3
($\hat{A}_{12}=0.92$, $p_{\mathrm{H}}=0.0045$) and libtiff
($\hat{A}_{12}=0.80$, $p_{\mathrm{H}}=0.040$) resolve after correction; the four
non-tied directions give exact sign $p=0.125$.

\subsection{RQ3: What each design needs}
\label{sec:eval:rq3}

\myparagraph{When the differentiating bugs arrive.}
The $25$ bugs triggered by every campaign of every scheduler have medians within
the first $11\%$ of the budget, with a median at $2\%$. Seven of Method~II's
differentiating bugs have medians from $5.6$ to $22.1$ hours, median $13.6$;
five arrive after the first quarter and XML001 in the final quarter
(Table~\ref{tab:ttb}; \S\ref{sec:appendix:arrival}).

\myparagraph{Domain-guided and learned context.}
Method~I's one phase signal gains $5.2\%$--$7.4\%$ over the baselines. Method~II
learns from \nContextFeatures\ signals and gains another $12.1\%$, leading
Method~I on seven targets and tying two. The environment varies sharply: nonzero
second-half reward ranges from $0.4\%$ of turns on libpng to $40.3\%$ on poppler,
and a model frozen after the first half fails to describe the second half in
$49.0\%$ of sqlite3 campaigns, $37.0\%$ of libxml2 campaigns, and $35.9\%$ of
poppler campaigns, versus below $10\%$ on every other target. The direct
comparison selects Method~II for \sys (\S\ref{sec:appendix:rq3}).

\myparagraph{Allocation learned during the campaign.}
No method parameter names a fuzzer or fixes its CPU share (goal~\textbf{G2}).
Method~II's largest share rises from $92.1\%$ to $96.6\%$ across quarters on
libtiff and from $37.6\%$ to $76.4\%$ on poppler, but stays near its initial
level on sqlite3 and libxml2. The target-dependent allocation follows the online
estimates. Method~I's concentration instead follows its first rotting verdict;
\base's largest share falls from $80.6\%$ to $24.5\%$ on libtiff and from
$74.7\%$ to $38.6\%$ on sqlite3
(Figure~\ref{fig:alloc}; \S\ref{sec:appendix:alloc}).

\subsection{RQ4: Unknown bugs}
\label{sec:eval:realworld}

RQ4 evaluates deployment on new targets (goal~\textbf{G3}). We run the
shipped Method~II configuration unchanged, without target-specific
pretraining, on \nRealPrograms C++ programs outside the Magma
evaluation. The harnesses invoke documented file-parsing entry points, and each
defines only its own \texttt{main} (\S\ref{sec:appendix:realworld}).

The unchanged deployment produces \nRealDefects previously unknown defect records:
$62$ in MNN, $42$ in Arm NN, $14$ in Krita, one in ncnn, and one in FreeCAD
(Table~\ref{tab:realworld}). We deduplicate by faulting function, file, line,
sanitizer-reported fault, and entry point after removing known and nonreproducing
candidates. Every record crashes the corresponding uninstrumented build in five
of five runs. MNN and Arm NN account for $104$ defects reached through model
conversion or deserialization, giving them a model-supply-chain exposure path
(\S\ref{sec:appendix:delivery}).

We reported every record through the channel prescribed by the project's security
policy. Triage places \nRealPublic in the crash class and \nRealPrivate in the
withheld class for memory corruption or information disclosure; $42$ crash-class
records have public tracker issues with working reproducers, while Arm NN's
remaining $31$ stay under vendor-requested coordination. The appendix reports the
available weakness, location, function, and disclosure details without publishing
a trigger or reproducing input. These results establish previously unknown,
reproducible crashing defects; independent exploitability is outside the evidence
evaluated here.

\section{Ablation}
\label{sec:ablation}

\begin{figure}[t]
\centering
\begin{tikzpicture}
\begin{axis}[
  width=0.78\columnwidth, height=4.3cm,
  xbar, bar width=6.5pt,
  xmin=-1.15, xmax=0.16,
  xtick={-1.0,-0.75,-0.5,-0.25,0},
  xlabel={change in mean unique bugs per target},
  xlabel style={font=\normalsize, yshift=3pt},
  symbolic y coords={ridge,models,discount,rank,form,context,explore},
  ytick=data,
  y=0.40cm,
  yticklabel style={font=\normalsize},
  xticklabel style={font=\normalsize},
  axis lines*=left,
  xmajorgrids, grid style={draw=black!10},
  tick align=outside,
  clip=false,
  nodes near coords, every node near coord/.append style={font=\normalsize, text=black!65},
  nodes near coords style={/pgf/number format/precision=2,
                           /pgf/number format/fixed,
                           /pgf/number format/fixed zerofill},
]
\addplot[draw=black!45, fill=black!12] coordinates
  {(0,ridge) (0,models) (-0.20,rank) (-0.20,form)};
\addplot[draw=cM2!80, fill=cM2!45] coordinates {(-0.69,context) (-1.00,explore)};
\draw[black!35, dashed, line width=0.5pt]
      (axis cs:0,ridge) -- (axis cs:0,explore);
\node[anchor=west, font=\normalsize, text=black!62] at (axis cs:0.01,discount)
      {held-out only};
\end{axis}
\end{tikzpicture}
\caption{Ablations of Method~II. Each bar reports the change in mean
unique bugs per target when one setting is replaced, with each pair run
at full campaign length under the protocol of \S\ref{sec:ablation}.
\emph{Explore} removes the Gaussian uncertainty term. \emph{Context}
replaces all \nContextFeatures\ signals with an intercept and therefore
also makes the context-derived confidence width identical across
fuzzers. \emph{Form} uses a linear instead of random-Fourier
representation; \emph{rank} uses a rank-3 projection; \emph{models}
shares one model across fuzzers; and \emph{ridge} increases $\ridge$
from 10 to 100. The colored bars mark context and exploration, the only
components whose removal produces a difference resolved by at least one
full-campaign pair. The remaining full-campaign replacements stay
within their pairs' resolution. Discounting is evaluated by held-out
reward prediction and therefore carries no bar
(\S\ref{sec:appendix:ablationfull}).}
\label{fig:ablation}
\end{figure}
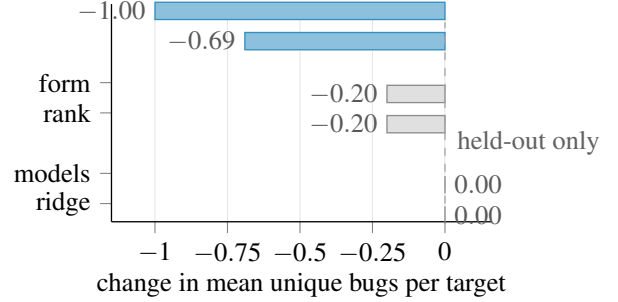

Goal~\textbf{G4} asks which components of Method~II produce its
full-campaign bug-count gain. Figure~\ref{fig:ablation} identifies two:
removing the context costs $0.69$ mean unique bugs per target, and
removing exploration costs $1.00$. None of the other full-campaign
alternatives in the figure produces a difference that its pair resolves.

\label{sec:ablation:judging}%
We evaluate changes to the deployed scheduler on full-campaign unique
bugs using the pairwise protocol of \S\ref{sec:eval:setup}. We evaluate
choices about predictive information, such as discounting past
observations, by leave-one-campaign-out held-out reward prediction.
\S\ref{sec:appendix:ablationfull} gives the per-pair measurements and
resolutions.

\myparagraph{The context.}
\label{sec:ablation:context}
We isolate the context-aware decision by replacing all
\nContextFeatures\ signals with an intercept while retaining the same
per-fuzzer ridge update, randomized selection rule, reward, substrate,
and constants. This intercept-only scheduler runs for the full campaign
on all \nTargets targets with \nTrials campaigns per cell. Because
\texttt{ctx\_unc} is one of the removed signals, the control also makes
the confidence width identical across fuzzers; it measures the complete
context-aware decision signal, including its context-derived uncertainty.

The intercept-only scheduler triggers $28.00$ mean unique bugs summed
across the suite, compared with $34.20$ for Method~II and $28.40$ for
\base. Restoring the complete context-aware decision therefore adds
$6.20$ bugs, while the same learning rule without that signal is
$0.40$ below \base. The gain over the intercept-only scheduler reaches
its pair-specific resolution on five targets: poppler at $+1.60$,
libtiff at $+1.40$, libxml2 at $+1.00$, and sqlite3 and libpng at
$+0.80$ each. The lua gap remains below its resolution; the pairs on
openssl, libsndfile, and php admit no resolved difference
(\S\ref{sec:appendix:ablationfull}).

Held-out prediction separates contextual information from the
uncertainty signal. A model using only an intercept and
\texttt{ctx\_unc} has $9.7\%$ more prediction error at the median fold
than the full context, performs worse on all \nTargets targets after
Holm correction, and performs better on none. Thus, the full-campaign
gain requires the complete context-aware decision, and the predictive
gain extends beyond its pull-count-derived uncertainty signal.

\myparagraph{The exploration term.}
\label{sec:ablation:settled}
Removing the Gaussian uncertainty term reduces summed mean unique bugs
from $34.20$ to $25.20$, a loss of $1.00$ per target, and reduces edge
coverage by $16\%$. This is the largest full-campaign ablation: without
exploration, Method~II trails every baseline and three of the
\nEngines fuzzers run alone.

The decision traces show the corresponding change in allocation. With
selection based only on the current predicted mean, no passed-over
fuzzer regains the highest prediction across any learned decision in
any campaign. The scheduler consequently locks onto one fuzzer, whereas
randomized uncertainty keeps fuzzers with weakly observed contexts in
contention.

Method~II uses a perturbation multiplier of one on every target.
Increasing $\ridge$ from 10 to 100 narrows the width by at most
$\sqrt{10}$ yet changes mean unique bugs by $0.00$ per target. The
result therefore persists across the measured scale change. The basis
width $\rff$, bandwidth $\sigma$, and multiplier itself were not varied;
the experiment establishes the contribution of exploration without
selecting an optimal scale (\S\ref{sec:appendix:select}).

\myparagraph{Remaining model choices.}
\label{sec:ablation:indistinguishable}
Increasing $\ridge$ from 10 to 100 and replacing the per-fuzzer models
with one shared model each changes mean unique bugs by $0.00$ per target.
A rank-3 projection and a linear representation each reduce the mean by
$0.20$; neither difference reaches its pair-specific resolution.

Held-out prediction determines the choices that full-campaign bug count
does not separate. No tested low-rank projection improves on the full
context on any target, and $34$ of the $54$ target-rank tests
significantly favor the full context. Both tested forgetting factors
increase prediction error on all \nTargets targets after Holm correction
and reduce it on none. At campaign-scale data volumes, the
random-Fourier representation has $6.9\%$--$36.1\%$ lower error than the
linear representation on both tested targets, and prediction also
favors separate per-fuzzer models.

The sweep does not vary $\rff$, $\sigma$, the one-turn warm-up, the
$\pm 5$ clamp, the 84-decision reward window, or the perturbation
multiplier. These remain fixed across targets; their roles are detailed
in \S\ref{sec:appendix:constants}.

\section{Discussion}
\label{sec:discussion}

\myparagraph{What to encode and what to learn.}
Goal~\textbf{G1} asks that general structure be encoded and
target-specific behavior be learned. Method~I encodes one such
structure---whether a fuzzer's reward is rising or rotting---and learns
the resulting allocation online. This domain guidance alone improves
summed mean unique bugs over the three baselines by $5.2\%$ to $7.4\%$.

Method~II instead learns the weighting of the full context during the
current campaign while treating the context-to-reward relationship as
fixed within that campaign. It improves on Method~I by another $12.1\%$,
leading on seven targets, tying on two, and trailing on none. This direct
comparison selects Method~II for \sys. The same configuration runs on
all \nTargets Magma targets and deploys unchanged on the
\nRealPrograms real-world programs, where it reveals \nRealDefects
previously unknown crashing defects.

These results place the design boundary: encode that a fuzzer's current
context and the uncertainty behind its prediction govern scheduling,
but learn how the context signals predict reward on the current target.

\myparagraph{Scope of the evidence.}
The comparative result covers \nTargets Magma targets,
\campaignHours-hour campaigns, and one pool of \nEngines fuzzers, four
of them AFL-lineage. Running every policy in the common dispatch loop
holds the harness, reward, seed sharing, resources, and campaign length
fixed, isolating the scheduling rule. The comparison therefore concerns
those rules under common conditions rather than end-to-end reproductions
of the released systems. An equal-share ensemble is not among the
comparators (\S\ref{sec:appendix:comparison}).

Proposition~\ref{prop:blind} establishes that context can be necessary;
it does not provide a regret guarantee for the deployed scheduler.
Method~II optimizes predicted next-turn reward under concurrent,
delayed, and coupled feedback. Its performance evidence is the observed
full-campaign bug count at the tested horizon; campaign-optimal
scheduling remains open.

The model-design analyses hold out campaigns within a target rather
than entire targets. The real-world evaluation establishes unchanged
deployment and validated findings rather than comparative gain. Finally,
the two motivating target statistics have rank correlations of $0.20$
and $0.00$ with target-level gain. The evaluation therefore establishes
the suite-level benefit of context but does not yet identify before a
campaign which targets will benefit most.

\section{Conclusion}
\label{sec:conclusion}

Ensemble scheduling should respond to the situation each fuzzer is in,
not only summarize what that fuzzer produced before. Across
$1{,}841$ campaign--fuzzer pairs, raw per-turn reward falls in $96.7\%$
of them, while consecutive-window rankings are no more stable than two
estimates formed within the same window. The useful weighting is also
target-specific: a model learned from eight targets predicts the ninth
worse than one learned within that target on eight of \nTargets targets,
and the two select different top-ranked fuzzers on a median $51.0\%$ of
decisions.

\sys turns these observations into two context-aware online-learning
methods. Method~I encodes the domain-guided distinction between rising
and rotting reward and improves summed mean unique bugs over the three
baselines by $5.2\%$ to $7.4\%$. Method~II learns how the full context
predicts reward during the current campaign. It improves on Method~I by
$12.1\%$ and triggers $20.4\%$, $18.8\%$, and $17.9\%$ more unique bugs
than the three baselines, or $11.8\%$ more than the strongest baseline
selected separately for each target. Its additional yield includes bugs
the baselines trigger rarely or never and continues into the final quarter.

The ablations identify the complete context-aware decision and
randomized uncertainty as the two components that produce the
full-campaign result: replacing the context with an intercept costs
$0.69$ mean unique bugs per target, and removing exploration costs
$1.00$. The same Method~II configuration deploys without
target-specific pretraining on \nRealPrograms widely used C++ programs,
where it finds \nRealDefects previously unknown, reproducible crashing
defects and reports all of them through the maintainers' prescribed
channels. Together, these results show that learning a target-specific
mapping from observable context to reward produces a stronger ensemble
than fixing how past fuzzer output is combined before the campaign.

\clearpage
\appendix
\section*{Ethical Considerations}
\label{sec:ethics}

This work identifies memory-safety defects in production software. We considered
four stakeholder groups: maintainers who triage and fix reports, users of the
affected software, the wider open-source community, and potential adversaries who
could use a defect disclosed before a patch exists.

\myparagraph{Disclosure.}
We reported each defect as soon as it was triaged and minimized, following the
security process published by the corresponding project. All \nRealDefects records
reproduce as crashes without sanitizers or coverage instrumentation. The worst
outcome observed across a defect's runs determined its disclosure class. Of the
\nRealPublic crash-class records, $42$ appear in public issue trackers with working
reproducers and $31$ remain under vendor-requested coordination. The
\nRealPrivate records involving memory corruption or information disclosure were
reported only through private security channels.

\myparagraph{Status snapshot.}
At the paper's status snapshot, no defect had been fixed and no CVE identifier had
been assigned. We did not request CVEs, following the recommendation that
maintainers do so~\cite{sok2024fuzzeval}. Eight public issues contain the $42$
publicly filed records across MNN, Krita, and ncnn, grouping defects that share
an entry point. Each issue records the program, version, weakness class, and
fault location also reported in
Tables~\ref{tab:locations} and~\ref{tab:locationsb}.

\myparagraph{Vendor coordination.}
One vendor requested in writing that public disclosure be coordinated with it.
Those records remain included in the totals of \S\ref{sec:eval:realworld} and
Table~\ref{tab:realworld}, but their locations, triggers, and reproducers do not
appear in the paper or artifact. They remain withheld until the vendor issues its
advisory.

\myparagraph{Publication boundary.}
The paper reports the affected program, version, and defect class for every record,
and source locations for records outside the coordinated vendor. It publishes no
trigger sequence or reproducing input. Public reproducers accompany the public
crash-class reports; memory-corruption, information-disclosure, and coordinated
vendor material remain withheld under the policy above. Every record was reported
to its maintainer immediately after triage.

\clearpage
\section*{Open Science}
\label{sec:openscience}
\addcontentsline{toc}{section}{Open Science}

The anonymous artifact is available at
\url{https://anonymous.4open.science/r/CORAL-fuzzing/README.md}. It contains the source,
configurations, campaign records, and analysis scripts needed to reproduce the
paper's results. The link performs no tracking. On acceptance, the records will
move to a stable public location under the authors' names.

\myparagraph{Artifact 1: the schedulers and the harness.} This artifact contains the
complete \sys source: both schedulers, the dispatch loop, all three baselines, adapters
for all \nEngines fuzzers, the shared seed store, and the evaluator. Container
definitions pin every fuzzer and target build, and the launcher reproduces a campaign
end to end. These materials rerun any cell of Table~\ref{tab:main}.

\myparagraph{Artifact 2: benchmark configuration and per-campaign data.} This artifact
contains per-target configuration files for every scheduler and fuzzer alone, including
every parameter value behind Table~\ref{tab:main}, and the per-campaign records behind
the reported results. Each record holds its configuration, its coverage and
unique-bug time series, and the Magma oracle's per-bug trigger record.
Table~\ref{tab:main}, Table~\ref{tab:perbug}, and Figure~\ref{fig:decay} are
regenerated from these records by artifact scripts. \texttt{results\_table.py}
prints the main and reward comparisons, \texttt{chat\_report.py} prints one
campaign's complete record, and the README gives the invocation for every
remaining table and figure.

\myparagraph{Artifact 3: the offline design analyses.} This artifact contains the
per-decision logs and analysis scripts behind the context and exploration ablations and
the setting sweep of \S\ref{sec:ablation}. They cover the held-out reward-prediction
comparison with and without the context, feature influence per feature, the
linear-against-non-linear measurement at every training volume, the rank and
leave-one-feature-out comparisons, the raw pre-scaling per-turn rate behind the decay
measurement of \S\ref{sec:motivation:decay}, the permutation test for coefficient
drift behind \S\ref{sec:eval:rq3}, and the frozen campaign list and
leave-one-campaign-out protocol they all read.

\myparagraph{Artifact 4: the real-world harnesses and reports.} This artifact contains
the harness source for each of the \nRealPrograms programs of
\S\ref{sec:eval:realworld}, the exact configuration from each build's cache, and the
per-defect reports. Each report gives the weakness class, full sanitizer output, and the
fault location for every defect outside the coordinated vendor. The public artifact
contains minimized crashing inputs for the $42$ publicly filed defects. Public reports
record the corresponding project and status. The anonymous-review artifact omits
direct tracker URLs; the post-review record will restore them as the authoritative
maintainer records. The artifact states which channel carried every defect.

\myparagraph{Withheld material and release conditions.}
The public artifact excludes minimized inputs for the \nRealPrivate records.
Crash-class reproducers follow the affected projects' public-reporting policies;
memory-corruption and information-disclosure reproducers remain restricted. The
classification uses the worst outcome observed across a defect's runs and was fixed
at triage before reporting. Restricted inputs will be released after the affected
project ships a fix or states that it will not act. If neither occurs, they remain
restricted. Material covered by the vendor's coordination request remains withheld
until that vendor authorizes release. The public record still contains the harness,
build configuration, symbolized fault location where permitted, and full sanitizer
output.

\myparagraph{Where it lives, and when it was counted.} The artifact is deposited in
an anonymous GitHub repository during review. The archival release will be deposited
as two stable records rather than depend on a source-hosting account. The public one
will carry the schedulers, harnesses, configurations, per-campaign data and the
complete defect table; the restricted one will carry the withheld reproducers. Every
status in Table~\ref{tab:realworld} is a snapshot at the time of writing. Statuses
change, and the projects' own records are authoritative over this paper's copy of them.

\myparagraph{A reduced protocol.} The artifact ships a reduced
protocol of one target and all five schedulers, at reduced repetition. It exercises
every code path the full evaluation does on a single workstation. The complete
per-campaign records ship with it, so the full table can be checked without
re-running it.

\clearpage
\bibliographystyle{plain}
\bibliography{refs}

\clearpage
\section{Supplementary Material}
\label{sec:appendix}

\subsection{The full context}
\label{sec:appendix:features}

\begin{table*}[t]
\centering
\tabfont
\caption{The \nContextFeatures\ context features, one vector per fuzzer per decision.
Every one is computed from quantities the dispatch loop already maintains. None requires a
bug oracle, a source annotation or a benchmark harness. Every window and code-property
feature is latched when a fuzzer's turn \emph{ends}, so nothing computed from a turn's
outcome depends on the turn whose reward the decision is predicting.
\texttt{time\_since\_run} and \texttt{elapsed\_frac} are clocks and advance while a fuzzer
waits, which is what makes the first a freshness signal.}
\label{tab:features}
\setlength{\tabcolsep}{3pt}
\begin{tabular}{@{}llp{13.5cm}@{}}
\toprule
& name & how it is computed \\
\midrule
\multicolumn{3}{@{}l}{\emph{recent productivity}} \\
& \texttt{win\_mean}            & mean reward over the window \\
& \texttt{win\_var}             & variance of the reward there \\
& \texttt{cov\_velocity}        & $1-e^{-\rho}$, $\rho$ the edges this fuzzer newly covered in the last $120$\,s of wall clock over the span of that window in seconds \\
& \texttt{slope}        & least-squares slope of the reward window \\
\midrule
\multicolumn{3}{@{}l}{\emph{trend and horizon}} \\
& \texttt{mk\_z}                & $\tanh(z)$, $z$ the Mann--Kendall statistic of the window standardized with the same continuity and tie corrections as \S\ref{sec:rr:phase} \\
& \texttt{horizon\_ratio}       & mean of the last 4 rewards over the window mean \\
& \texttt{horizon2\_ratio}      & the same over the last 2 \\
& \texttt{horizon8\_ratio}      & the same over the last 8 \\
\midrule
\multicolumn{3}{@{}l}{\emph{plateau and freshness}} \\
& \texttt{rounds\_since\_improve} & rounds since it last added an edge, over the window length \\
& \texttt{time\_since\_run}     & time since this fuzzer last ran, over the campaign so far \\
& \texttt{elapsed\_frac}        & fraction of the campaign budget consumed \\
\midrule
\multicolumn{3}{@{}l}{\emph{what the covered code looks like}} \\
& \texttt{g\_rarity}            & exponential moving average, rate $0.3$, over completed turns of the mean of $1/\log(2 + h_e)$ across that turn's newly covered edges, $h_e$ the ensemble-wide hit count of edge $e$ at that time \\
& \texttt{g\_sec}               & exponential moving average, rate $0.3$, of $1-e^{-\bar{s}}$ over completed turns, $\bar{s}$ the mean over that turn's newly covered edges of $s_e$, the count of memory-handling calls the block entered by $e$ makes, read once from the instrumented binary at build time~\cite{wang2020tortoisefuzz}; no source annotation and no bug oracle \\
& \texttt{g\_bug}       & exponential moving average, rate $0.3$, over completed turns of $1-e^{-c}$, $c$ the number of this fuzzer's own inputs that crashed in that turn; a crash count, not a bug count, and no oracle reads it \\
\midrule
\multicolumn{3}{@{}l}{\emph{estimate quality}} \\
& \texttt{ctx\_unc}             & $1/\sqrt{1+n_\arm}$, $n_\arm$ this fuzzer's turns so far \\
\bottomrule
\end{tabular}
\end{table*}

\subsection{Per-target bug ceilings}
\label{sec:appendix:ceilings}

\begin{table}[t]
\centering
\tabfont
\caption{What was available to be found. \emph{injected} is the number of bugs
Magma re-introduces into the target; \emph{reachable} is how many of those the
Magma oracle records as reached through the single harness we fuzz. The second
column bounds every number in Table~\ref{tab:main}.}
\label{tab:ceilings}
\setlength{\tabcolsep}{3pt}
\begin{tabular}{@{}llrr@{}}
\toprule
target & harness & injected & reachable \\
\midrule
libpng     & \texttt{libpng\_read} &  7 &  6 \\
libsndfile & \texttt{sndfile}      & 18 &  8 \\
libtiff    & \texttt{tiff\_rgba}     & 14 &  9 \\
libxml2    & \texttt{xml\_reader} & 17 & 10 \\
lua        & \texttt{lua}                  &  4 &  4 \\
openssl    & \texttt{asn1}                 & 20 &  4 \\
php        & \texttt{exif}                 & 16 &  5 \\
poppler    & \texttt{pdf}          & 22 & 17 \\
sqlite3    & \texttt{sqlite3}        & 20 & 14 \\
\bottomrule
\end{tabular}
\end{table}

\subsection{Time to trigger each bug}
\label{sec:appendix:ttb}

\begin{table}[t]
\centering
\tabfont
\caption{Time to trigger, over every bug the context-aware scheduler and the baselines do not trigger equally often, hardest first. \emph{baseline} is the strongest of the three on that bug, taken as the one that triggers it in the most campaigns and then the one that reaches its median soonest. \emph{camp} is how many of that cell's \nTrials campaigns trigger the bug with a sampled trigger time, \emph{med} the Kaplan--Meier median, \emph{---} that fewer than half trigger it so no median falls in budget.}
\label{tab:ttb}
\setlength{\tabcolsep}{3.5pt}
\begin{tabular}{@{}llrrrr@{}}
\toprule
& & \multicolumn{2}{c}{baseline} & \multicolumn{2}{c}{\ctxshort} \\
\cmidrule(lr){3-4}\cmidrule(lr){5-6}
target & bug & camp & med & camp & med \\
\midrule
sqlite3 & SQL013 & 0 & --- & \textbf{3} & --- \\
poppler & PDF021 & 0 & --- & \textbf{1} & --- \\
sqlite3 & SQL015 & 0 & --- & \textbf{1} & --- \\
libtiff & TIF001 & 1 & --- & \textbf{3} & --- \\
sqlite3 & SQL012 & 1 & --- & \textbf{3} & --- \\
poppler & PDF008 & 1 & --- & 0 & --- \\
libtiff & TIF008 & 2 & --- & \textbf{8} & \textbf{11.6\,h} \\
libxml2 & XML001 & 2 & --- & \textbf{5} & \textbf{22.1\,h} \\
lua & LUA001 & 2 & --- & \textbf{3} & --- \\
lua & LUA003 & 2 & --- & \textbf{3} & --- \\
sqlite3 & SQL020 & 3 & --- & \textbf{5} & \textbf{15.1\,h} \\
libxml2 & XML009 & 4 & --- & \textbf{6} & \textbf{13.7\,h} \\
libpng & PNG001 & 6 & 18.7\,h & \textbf{10} & 5.8\,h \\
poppler & PDF012 & 6 & 10.8\,h & \textbf{8} & 5.6\,h \\
poppler & PDF018 & 8 & 11.2\,h & \textbf{10} & 13.6\,h \\
libtiff & TIF002 & 10 & 10.6\,h & 9 & 10.1\,h \\
\bottomrule
\end{tabular}
\end{table}

\subsection{Real-world weakness classes}
\label{sec:appendix:weakness}

\begin{table}[t]
\centering
\tabfont
\caption{Weakness classes of the \nRealDefects real-world defects, taken from the
class each defect document assigns from its sanitizer evidence and the source at the
fault site. Five reports carry no class of their own and are counted as such.}
\label{tab:weakness}
\setlength{\tabcolsep}{4pt}
\begin{tabular}{@{}llr@{}}
\toprule
class & weakness & defects \\
\midrule
CWE-476 & null pointer dereference                     & 54 \\
CWE-125 & out-of-bounds read                           & 34 \\
CWE-129 & improper validation of an array index        & 11 \\
CWE-787 & out-of-bounds write                          &  6 \\
CWE-369 & divide by zero                               &  3 \\
CWE-789 & memory allocation with an excessive size     &  3 \\
\multicolumn{2}{@{}l}{other (integer overflow, integer underflow,} &  \\
\multicolumn{2}{@{}l}{\quad type confusion, use after free)} &  4 \\
\multicolumn{2}{@{}l}{no class in the report} &  5 \\
\midrule
\multicolumn{2}{@{}l}{\emph{total}} & \bestcell{\nRealDefects} \\
\bottomrule
\end{tabular}
\end{table}

\subsection{Where each defect is}
\label{sec:appendix:locations}

\begin{table*}[t]
\centering
\scriptsize
\caption{Where each defect is, part 1 of 2. Each row gives the class its report carries, the file and line the sanitizer names, the function where it names one, the fault it reports, and the entry point that reached it. A dash in the class column marks a defect whose report carries no class, and in the entry column one whose trace does not name the format being parsed. Two rows carry a file without a line, function or fault, because the sanitizer attributes the fault to a translation unit rather than a statement; both are counted in Table~\ref{tab:weakness} under the class their report carries. The last four fields are the deduplication rule of \S\ref{sec:eval:realworld}, and every row differs from every other in at least one of them. We print no trigger and no reproducing input.}
\label{tab:locations}
\setlength{\tabcolsep}{4pt}
\begin{tabular}{@{}llllll@{}}
\toprule
program & CWE & file\,:\,line & function & fault & entry \\
\midrule
MNN & 125 & \texttt{\fontsize{6.4}{7.6}\selectfont liteConverter.cpp:421} & \texttt{\fontsize{6.4}{7.6}\selectfont readModel} & {\fontsize{6.4}{7.6}\selectfont wild pointer} & {\fontsize{6.4}{7.6}\selectfont tflite} \\
 & 125 & \texttt{\fontsize{6.4}{7.6}\selectfont liteConverter.cpp:421} & \texttt{\fontsize{6.4}{7.6}\selectfont readModel} & {\fontsize{6.4}{7.6}\selectfont heap read} & {\fontsize{6.4}{7.6}\selectfont tflite} \\
 & 476 & \texttt{\fontsize{6.4}{7.6}\selectfont Program.cpp:122} & \texttt{\fontsize{6.4}{7.6}\selectfont create} & {\fontsize{6.4}{7.6}\selectfont wild pointer} & {\fontsize{6.4}{7.6}\selectfont --} \\
 & 125 & \texttt{\fontsize{6.4}{7.6}\selectfont liteConverter.cpp:209} & \texttt{\fontsize{6.4}{7.6}\selectfont tflite2MNNNet} & {\fontsize{6.4}{7.6}\selectfont heap read} & {\fontsize{6.4}{7.6}\selectfont tflite} \\
 & 476 & \texttt{\fontsize{6.4}{7.6}\selectfont liteConverter.cpp:235} & \texttt{\fontsize{6.4}{7.6}\selectfont tflite2MNNNet} & {\fontsize{6.4}{7.6}\selectfont wild pointer} & {\fontsize{6.4}{7.6}\selectfont tflite} \\
 & 125 & \texttt{\fontsize{6.4}{7.6}\selectfont liteConverter.cpp:264} & \texttt{\fontsize{6.4}{7.6}\selectfont tflite2MNNNet} & {\fontsize{6.4}{7.6}\selectfont wild pointer} & {\fontsize{6.4}{7.6}\selectfont tflite} \\
 & 125 & \texttt{\fontsize{6.4}{7.6}\selectfont liteConverter.cpp:269} & \texttt{\fontsize{6.4}{7.6}\selectfont tflite2MNNNet} & {\fontsize{6.4}{7.6}\selectfont heap read} & {\fontsize{6.4}{7.6}\selectfont tflite} \\
 & 476 & \texttt{\fontsize{6.4}{7.6}\selectfont liteConverter.cpp:344} & \texttt{\fontsize{6.4}{7.6}\selectfont tflite2MNNNet} & {\fontsize{6.4}{7.6}\selectfont wild pointer} & {\fontsize{6.4}{7.6}\selectfont tflite} \\
 & 125 & \texttt{\fontsize{6.4}{7.6}\selectfont Expr.hpp:52} & \texttt{\fontsize{6.4}{7.6}\selectfont VARP} & {\fontsize{6.4}{7.6}\selectfont heap read} & {\fontsize{6.4}{7.6}\selectfont tflite} \\
 & 476 & \texttt{\fontsize{6.4}{7.6}\selectfont ConvolutionTflite.cpp:59} & \texttt{\fontsize{6.4}{7.6}\selectfont run} & {\fontsize{6.4}{7.6}\selectfont wild pointer} & {\fontsize{6.4}{7.6}\selectfont tflite} \\
 & 129 & \texttt{\fontsize{6.4}{7.6}\selectfont ConvolutionTflite.cpp:64} & \texttt{\fontsize{6.4}{7.6}\selectfont run} & {\fontsize{6.4}{7.6}\selectfont wild pointer} & {\fontsize{6.4}{7.6}\selectfont tflite} \\
 & 369 & \texttt{\fontsize{6.4}{7.6}\selectfont ConvolutionTflite.cpp:70} & \texttt{\fontsize{6.4}{7.6}\selectfont run} & {\fontsize{6.4}{7.6}\selectfont arith} & {\fontsize{6.4}{7.6}\selectfont tflite} \\
 & 476 & \texttt{\fontsize{6.4}{7.6}\selectfont ConvolutionTflite.cpp:272} & \texttt{\fontsize{6.4}{7.6}\selectfont run} & {\fontsize{6.4}{7.6}\selectfont wild pointer} & {\fontsize{6.4}{7.6}\selectfont tflite} \\
 & 129 & \texttt{\fontsize{6.4}{7.6}\selectfont ConvolutionTflite.cpp:298} & \texttt{\fontsize{6.4}{7.6}\selectfont run} & {\fontsize{6.4}{7.6}\selectfont wild pointer} & {\fontsize{6.4}{7.6}\selectfont tflite} \\
 & 129 & \texttt{\fontsize{6.4}{7.6}\selectfont ConvolutionTflite.cpp:326} & \texttt{\fontsize{6.4}{7.6}\selectfont run} & {\fontsize{6.4}{7.6}\selectfont heap read} & {\fontsize{6.4}{7.6}\selectfont tflite} \\
 & 476 & \texttt{\fontsize{6.4}{7.6}\selectfont ConvolutionTflite.cpp:327} & \texttt{\fontsize{6.4}{7.6}\selectfont run} & {\fontsize{6.4}{7.6}\selectfont wild pointer} & {\fontsize{6.4}{7.6}\selectfont tflite} \\
 & 476 & \texttt{\fontsize{6.4}{7.6}\selectfont DepthwiseConv2DTflite.cpp:32} & \texttt{\fontsize{6.4}{7.6}\selectfont \_writeCommon} & {\fontsize{6.4}{7.6}\selectfont heap read} & {\fontsize{6.4}{7.6}\selectfont tflite} \\
 & 787 & \texttt{\fontsize{6.4}{7.6}\selectfont DepthwiseConv2DTflite.cpp:222} & \texttt{\fontsize{6.4}{7.6}\selectfont run} & {\fontsize{6.4}{7.6}\selectfont heap read} & {\fontsize{6.4}{7.6}\selectfont tflite} \\
 & 476 & \texttt{\fontsize{6.4}{7.6}\selectfont ConcatTflite.cpp:25} & \texttt{\fontsize{6.4}{7.6}\selectfont run} & {\fontsize{6.4}{7.6}\selectfont heap read} & {\fontsize{6.4}{7.6}\selectfont tflite} \\
 & 476 & \texttt{\fontsize{6.4}{7.6}\selectfont PoolingTflite.cpp:74} & \texttt{\fontsize{6.4}{7.6}\selectfont run} & {\fontsize{6.4}{7.6}\selectfont wild pointer} & {\fontsize{6.4}{7.6}\selectfont tflite} \\
 & 125 & \texttt{\fontsize{6.4}{7.6}\selectfont ResizeBilinear.cpp:55} & \texttt{\fontsize{6.4}{7.6}\selectfont run} & {\fontsize{6.4}{7.6}\selectfont wild pointer} & {\fontsize{6.4}{7.6}\selectfont tflite} \\
 & 476 & \texttt{\fontsize{6.4}{7.6}\selectfont TfliteUtils.cpp:107} & \texttt{\fontsize{6.4}{7.6}\selectfont convertDataFormatTflite} & {\fontsize{6.4}{7.6}\selectfont wild pointer} & {\fontsize{6.4}{7.6}\selectfont tflite} \\
 & 476 & \texttt{\fontsize{6.4}{7.6}\selectfont liteOpConverter.cpp:11} & \texttt{\fontsize{6.4}{7.6}\selectfont getOpCode} & {\fontsize{6.4}{7.6}\selectfont wild pointer} & {\fontsize{6.4}{7.6}\selectfont tflite} \\
 & 129 & \texttt{\fontsize{6.4}{7.6}\selectfont liteConverter.cpp:212} & \texttt{\fontsize{6.4}{7.6}\selectfont tflite2MNNNet} & {\fontsize{6.4}{7.6}\selectfont heap read} & {\fontsize{6.4}{7.6}\selectfont tflite} \\
 & 129 & \texttt{\fontsize{6.4}{7.6}\selectfont liteConverter.cpp:214} & \texttt{\fontsize{6.4}{7.6}\selectfont tflite2MNNNet} & {\fontsize{6.4}{7.6}\selectfont wild pointer} & {\fontsize{6.4}{7.6}\selectfont tflite} \\
 & 129 & \texttt{\fontsize{6.4}{7.6}\selectfont liteConverter.cpp:250} & \texttt{\fontsize{6.4}{7.6}\selectfont tflite2MNNNet} & {\fontsize{6.4}{7.6}\selectfont heap read} & {\fontsize{6.4}{7.6}\selectfont tflite} \\
 & 129 & \texttt{\fontsize{6.4}{7.6}\selectfont liteConverter.cpp:270} & \texttt{\fontsize{6.4}{7.6}\selectfont tflite2MNNNet} & {\fontsize{6.4}{7.6}\selectfont wild pointer} & {\fontsize{6.4}{7.6}\selectfont tflite} \\
 & 787 & \texttt{\fontsize{6.4}{7.6}\selectfont liteConverter.cpp:319} & \texttt{\fontsize{6.4}{7.6}\selectfont tflite2MNNNet} & {\fontsize{6.4}{7.6}\selectfont heap read} & {\fontsize{6.4}{7.6}\selectfont tflite} \\
 & 476 & \texttt{\fontsize{6.4}{7.6}\selectfont ConvolutionTflite.cpp:451} & \texttt{\fontsize{6.4}{7.6}\selectfont run} & {\fontsize{6.4}{7.6}\selectfont wild pointer} & {\fontsize{6.4}{7.6}\selectfont tflite} \\
 & 787 & \texttt{\fontsize{6.4}{7.6}\selectfont MoveUnaryOpBeforeReshape.cpp:59} & \texttt{\fontsize{6.4}{7.6}\selectfont onExecute} & {\fontsize{6.4}{7.6}\selectfont heap read} & {\fontsize{6.4}{7.6}\selectfont tflite} \\
 & 787 & \texttt{\fontsize{6.4}{7.6}\selectfont MoveUnaryOpBeforeReshape.cpp:66} & \texttt{\fontsize{6.4}{7.6}\selectfont onExecute} & {\fontsize{6.4}{7.6}\selectfont heap read} & {\fontsize{6.4}{7.6}\selectfont --} \\
 & 369 & \texttt{\fontsize{6.4}{7.6}\selectfont TransformGroupConvolution.cpp:191} & \texttt{\fontsize{6.4}{7.6}\selectfont onExecute} & {\fontsize{6.4}{7.6}\selectfont arith} & {\fontsize{6.4}{7.6}\selectfont --} \\
 & 476 & \texttt{\fontsize{6.4}{7.6}\selectfont PostConverter.cpp:713} & \texttt{\fontsize{6.4}{7.6}\selectfont optimizeNet} & {\fontsize{6.4}{7.6}\selectfont wild pointer} & {\fontsize{6.4}{7.6}\selectfont --} \\
 & 476 & \texttt{\fontsize{6.4}{7.6}\selectfont Expr.hpp:52} & \texttt{\fontsize{6.4}{7.6}\selectfont VARP} & {\fontsize{6.4}{7.6}\selectfont wild pointer} & {\fontsize{6.4}{7.6}\selectfont tflite} \\
 & 476 & \texttt{\fontsize{6.4}{7.6}\selectfont Expr.cpp:1287} & \texttt{\fontsize{6.4}{7.6}\selectfont save} & {\fontsize{6.4}{7.6}\selectfont wild pointer} & {\fontsize{6.4}{7.6}\selectfont --} \\
 & 476 & \texttt{\fontsize{6.4}{7.6}\selectfont Executor.cpp} & \texttt{\fontsize{6.4}{7.6}\selectfont } & {\fontsize{6.4}{7.6}\selectfont } & {\fontsize{6.4}{7.6}\selectfont --} \\
 & 787 & \texttt{\fontsize{6.4}{7.6}\selectfont Tensor.hpp:275} & \texttt{\fontsize{6.4}{7.6}\selectfont setLength} & {\fontsize{6.4}{7.6}\selectfont heap read} & {\fontsize{6.4}{7.6}\selectfont --} \\
 & 416 & \texttt{\fontsize{6.4}{7.6}\selectfont TensorUtils.hpp:86} & \texttt{\fontsize{6.4}{7.6}\selectfont ~NativeInsideDescribe} & {\fontsize{6.4}{7.6}\selectfont wild pointer} & {\fontsize{6.4}{7.6}\selectfont --} \\
 & 190 & \texttt{\fontsize{6.4}{7.6}\selectfont MNNMemoryUtils.cpp:24} & \texttt{\fontsize{6.4}{7.6}\selectfont MNNMemoryAllocAlign} & {\fontsize{6.4}{7.6}\selectfont allocation of } & {\fontsize{6.4}{7.6}\selectfont --} \\
\bottomrule
\end{tabular}
\end{table*}

\begin{table*}[t]
\centering
\scriptsize
\caption{Where each defect is, part 2 of 2. Each row gives the class its report carries, the file and line the sanitizer names, the function where it names one, the fault it reports, and the entry point that reached it. A dash in the class column marks a defect whose report carries no class, and in the entry column one whose trace does not name the format being parsed. Two rows carry a file without a line, function or fault, because the sanitizer attributes the fault to a translation unit rather than a statement; both are counted in Table~\ref{tab:weakness} under the class their report carries. The last four fields are the deduplication rule of \S\ref{sec:eval:realworld}, and every row differs from every other in at least one of them. We print no trigger and no reproducing input. The 42 defects in Arm NN are absent because the report we sent that vendor undertakes to publish nothing about them until they agree, and a row here would be publishing.}
\label{tab:locationsb}
\setlength{\tabcolsep}{4pt}
\begin{tabular}{@{}llllll@{}}
\toprule
program & CWE & file\,:\,line & function & fault & entry \\
\midrule
 & 129 & \texttt{\fontsize{6.4}{7.6}\selectfont ShapeConcat.cpp:62} & \texttt{\fontsize{6.4}{7.6}\selectfont onComputeSize} & {\fontsize{6.4}{7.6}\selectfont wild pointer} & {\fontsize{6.4}{7.6}\selectfont --} \\
 & 129 & \texttt{\fontsize{6.4}{7.6}\selectfont DepthwiseConv2DTflite.cpp:83} & \texttt{\fontsize{6.4}{7.6}\selectfont run} & {\fontsize{6.4}{7.6}\selectfont wild pointer} & {\fontsize{6.4}{7.6}\selectfont tflite} \\
 & 843 & \texttt{\fontsize{6.4}{7.6}\selectfont CustomTflite.cpp:111} & \texttt{\fontsize{6.4}{7.6}\selectfont run} & {\fontsize{6.4}{7.6}\selectfont wild pointer} & {\fontsize{6.4}{7.6}\selectfont tflite} \\
 & -- & \texttt{\fontsize{6.4}{7.6}\selectfont ConvolutionFloatFactory.cpp:269} & \texttt{\fontsize{6.4}{7.6}\selectfont create} & {\fontsize{6.4}{7.6}\selectfont arith} & {\fontsize{6.4}{7.6}\selectfont onnx} \\
 & 129 & \texttt{\fontsize{6.4}{7.6}\selectfont FuseDupOp.cpp:122} & \texttt{\fontsize{6.4}{7.6}\selectfont onExecute} & {\fontsize{6.4}{7.6}\selectfont wild pointer} & {\fontsize{6.4}{7.6}\selectfont tflite} \\
 & -- & \texttt{\fontsize{6.4}{7.6}\selectfont ShapeConvolution.cpp:86} & \texttt{\fontsize{6.4}{7.6}\selectfont onComputeSize} & {\fontsize{6.4}{7.6}\selectfont arith} & {\fontsize{6.4}{7.6}\selectfont onnx} \\
 & -- & \texttt{\fontsize{6.4}{7.6}\selectfont ConvolutionTiledExecutor.cpp:170} & \texttt{\fontsize{6.4}{7.6}\selectfont turnIm2ColToBlitInfo} & {\fontsize{6.4}{7.6}\selectfont arith} & {\fontsize{6.4}{7.6}\selectfont onnx} \\
 & 125 & \texttt{\fontsize{6.4}{7.6}\selectfont BufferAllocator.cpp:538} & \texttt{\fontsize{6.4}{7.6}\selectfont free} & {\fontsize{6.4}{7.6}\selectfont use after free} & {\fontsize{6.4}{7.6}\selectfont onnx} \\
 & 125 & \texttt{\fontsize{6.4}{7.6}\selectfont OnnxExtraManager.cpp:57} & \texttt{\fontsize{6.4}{7.6}\selectfont operator} & {\fontsize{6.4}{7.6}\selectfont wild pointer} & {\fontsize{6.4}{7.6}\selectfont onnx} \\
 & 476 & \texttt{\fontsize{6.4}{7.6}\selectfont Expr.cpp:240} & \texttt{\fontsize{6.4}{7.6}\selectfont create} & {\fontsize{6.4}{7.6}\selectfont wild pointer} & {\fontsize{6.4}{7.6}\selectfont onnx} \\
 & 125 & \texttt{\fontsize{6.4}{7.6}\selectfont OnnxConvolutionMerge.cpp:328} & \texttt{\fontsize{6.4}{7.6}\selectfont onExecute} & {\fontsize{6.4}{7.6}\selectfont heap read} & {\fontsize{6.4}{7.6}\selectfont onnx} \\
 & 125 & \texttt{\fontsize{6.4}{7.6}\selectfont OnnxPooling.cpp:76} & \texttt{\fontsize{6.4}{7.6}\selectfont onExecute} & {\fontsize{6.4}{7.6}\selectfont wild pointer} & {\fontsize{6.4}{7.6}\selectfont onnx} \\
 & 476 & \texttt{\fontsize{6.4}{7.6}\selectfont Expr.hpp:52} & \texttt{\fontsize{6.4}{7.6}\selectfont VARP} & {\fontsize{6.4}{7.6}\selectfont wild pointer} & {\fontsize{6.4}{7.6}\selectfont onnx} \\
 & 476 & \texttt{\fontsize{6.4}{7.6}\selectfont Expr.hpp:123} & \texttt{\fontsize{6.4}{7.6}\selectfont expr} & {\fontsize{6.4}{7.6}\selectfont wild pointer} & {\fontsize{6.4}{7.6}\selectfont onnx} \\
 & -- & \texttt{\fontsize{6.4}{7.6}\selectfont ShapePool.cpp:74} & \texttt{\fontsize{6.4}{7.6}\selectfont onComputeSize} & {\fontsize{6.4}{7.6}\selectfont arith} & {\fontsize{6.4}{7.6}\selectfont onnx} \\
 & -- & \texttt{\fontsize{6.4}{7.6}\selectfont TransformGroupConvolution.cpp:191} & \texttt{\fontsize{6.4}{7.6}\selectfont onExecute} & {\fontsize{6.4}{7.6}\selectfont arith} & {\fontsize{6.4}{7.6}\selectfont onnx} \\
 & 125 & \texttt{\fontsize{6.4}{7.6}\selectfont GemmAVX2FMA.cpp:73} & \texttt{\fontsize{6.4}{7.6}\selectfont \_AVX\_MNNComputeMatMulForE\_1FMA} & {\fontsize{6.4}{7.6}\selectfont heap read} & {\fontsize{6.4}{7.6}\selectfont onnx} \\
 & 476 & \texttt{\fontsize{6.4}{7.6}\selectfont repeated\_ptr\_field.h:857} & \texttt{\fontsize{6.4}{7.6}\selectfont TypeHandler>} & {\fontsize{6.4}{7.6}\selectfont wild pointer} & {\fontsize{6.4}{7.6}\selectfont onnx} \\
 & 125 & \texttt{\fontsize{6.4}{7.6}\selectfont onnxConverter.cpp} & \texttt{\fontsize{6.4}{7.6}\selectfont } & {\fontsize{6.4}{7.6}\selectfont } & {\fontsize{6.4}{7.6}\selectfont onnx} \\
 & 125 & \texttt{\fontsize{6.4}{7.6}\selectfont onnxOpConverter.cpp:333} & \texttt{\fontsize{6.4}{7.6}\selectfont convertTensorToBlob} & {\fontsize{6.4}{7.6}\selectfont heap read} & {\fontsize{6.4}{7.6}\selectfont onnx} \\
 & 125 & \texttt{\fontsize{6.4}{7.6}\selectfont onnxOpConverter.cpp:342} & \texttt{\fontsize{6.4}{7.6}\selectfont convertTensorToBlob} & {\fontsize{6.4}{7.6}\selectfont heap read} & {\fontsize{6.4}{7.6}\selectfont onnx} \\
 & 125 & \texttt{\fontsize{6.4}{7.6}\selectfont onnxOpConverter.cpp:358} & \texttt{\fontsize{6.4}{7.6}\selectfont convertTensorToBlob} & {\fontsize{6.4}{7.6}\selectfont heap read} & {\fontsize{6.4}{7.6}\selectfont onnx} \\
 & 125 & \texttt{\fontsize{6.4}{7.6}\selectfont onnxOpConverter.cpp:412} & \texttt{\fontsize{6.4}{7.6}\selectfont convertTensorToBlob} & {\fontsize{6.4}{7.6}\selectfont heap read} & {\fontsize{6.4}{7.6}\selectfont onnx} \\
Krita & 125 & \texttt{\fontsize{6.4}{7.6}\selectfont kis\_gbr\_brush.cpp:236} & \texttt{\fontsize{6.4}{7.6}\selectfont init} & {\fontsize{6.4}{7.6}\selectfont heap read} & {\fontsize{6.4}{7.6}\selectfont --} \\
 & 125 & \texttt{\fontsize{6.4}{7.6}\selectfont kis\_gbr\_brush.cpp:256} & \texttt{\fontsize{6.4}{7.6}\selectfont init} & {\fontsize{6.4}{7.6}\selectfont heap read} & {\fontsize{6.4}{7.6}\selectfont --} \\
 & 789 & \texttt{\fontsize{6.4}{7.6}\selectfont kis\_gbr\_brush.cpp:188} & \texttt{\fontsize{6.4}{7.6}\selectfont init} & {\fontsize{6.4}{7.6}\selectfont allocation of } & {\fontsize{6.4}{7.6}\selectfont --} \\
 & 125 & \texttt{\fontsize{6.4}{7.6}\selectfont kis\_gbr\_brush.cpp:141} & \texttt{\fontsize{6.4}{7.6}\selectfont init} & {\fontsize{6.4}{7.6}\selectfont wild pointer} & {\fontsize{6.4}{7.6}\selectfont --} \\
 & 125 & \texttt{\fontsize{6.4}{7.6}\selectfont kis\_abr\_brush\_collection.cpp:47} & \texttt{\fontsize{6.4}{7.6}\selectfont convertToQImage} & {\fontsize{6.4}{7.6}\selectfont heap read} & {\fontsize{6.4}{7.6}\selectfont --} \\
 & 476 & \texttt{\fontsize{6.4}{7.6}\selectfont kis\_abr\_brush\_collection.cpp:48} & \texttt{\fontsize{6.4}{7.6}\selectfont convertToQImage} & {\fontsize{6.4}{7.6}\selectfont wild pointer} & {\fontsize{6.4}{7.6}\selectfont --} \\
 & 787 & \texttt{\fontsize{6.4}{7.6}\selectfont kis\_abr\_brush\_collection.cpp:95} & \texttt{\fontsize{6.4}{7.6}\selectfont rle\_decode} & {\fontsize{6.4}{7.6}\selectfont heap read} & {\fontsize{6.4}{7.6}\selectfont --} \\
 & 476 & \texttt{\fontsize{6.4}{7.6}\selectfont kis\_asl\_reader.cpp:427} & \texttt{\fontsize{6.4}{7.6}\selectfont readVirtualArrayList<} & {\fontsize{6.4}{7.6}\selectfont wild pointer} & {\fontsize{6.4}{7.6}\selectfont --} \\
 & 125 & \texttt{\fontsize{6.4}{7.6}\selectfont kis\_asl\_reader.cpp:561} & \texttt{\fontsize{6.4}{7.6}\selectfont readPattern<} & {\fontsize{6.4}{7.6}\selectfont wild pointer} & {\fontsize{6.4}{7.6}\selectfont --} \\
 & 476 & \texttt{\fontsize{6.4}{7.6}\selectfont kis\_brushes\_pipe.h:47} & \texttt{\fontsize{6.4}{7.6}\selectfont lastBrush} & {\fontsize{6.4}{7.6}\selectfont wild pointer} & {\fontsize{6.4}{7.6}\selectfont --} \\
 & 789 & \texttt{\fontsize{6.4}{7.6}\selectfont compression.cpp:90} & \texttt{\fontsize{6.4}{7.6}\selectfont decompress} & {\fontsize{6.4}{7.6}\selectfont allocation of } & {\fontsize{6.4}{7.6}\selectfont --} \\
 & 789 & \texttt{\fontsize{6.4}{7.6}\selectfont compression.cpp:367} & \texttt{\fontsize{6.4}{7.6}\selectfont decompress} & {\fontsize{6.4}{7.6}\selectfont allocation of } & {\fontsize{6.4}{7.6}\selectfont --} \\
 & 125 & \texttt{\fontsize{6.4}{7.6}\selectfont compression.cpp:258} & \texttt{\fontsize{6.4}{7.6}\selectfont psd\_unzip\_with\_prediction<unsigned char>} & {\fontsize{6.4}{7.6}\selectfont heap read} & {\fontsize{6.4}{7.6}\selectfont --} \\
 & 125 & \texttt{\fontsize{6.4}{7.6}\selectfont compression.cpp:276} & \texttt{\fontsize{6.4}{7.6}\selectfont psd\_unzip\_with\_prediction<unsigned short>} & {\fontsize{6.4}{7.6}\selectfont heap read} & {\fontsize{6.4}{7.6}\selectfont --} \\
ncnn & 369 & \texttt{\fontsize{6.4}{7.6}\selectfont multiheadattention.cpp:59} & \texttt{\fontsize{6.4}{7.6}\selectfont load\_param} & {\fontsize{6.4}{7.6}\selectfont arith} & {\fontsize{6.4}{7.6}\selectfont --} \\
FreeCAD & 191 & \texttt{\fontsize{6.4}{7.6}\selectfont MeshKernel.cpp:396} & \texttt{\fontsize{6.4}{7.6}\selectfont Merge} & {\fontsize{6.4}{7.6}\selectfont heap read} & {\fontsize{6.4}{7.6}\selectfont --} \\
\bottomrule
\end{tabular}
\end{table*}

\subsection{Alternative reward definitions}
\label{sec:appendix:reward}

\begin{table*}[t]
\centering
\tabfont
\caption{The reward the scheduler optimizes, swept on the \ctx.
\texttt{coverage\_interval} is \base's definition and the one every other result in this
paper uses. \texttt{frontier\_branch} credits a turn for branches at the edge of the
shared map, and \texttt{rarity\_density} for the rarity of the edges it covers. Mean
unique bugs over \nTrials campaigns per cell, read against the context-aware row of
Table~\ref{tab:main}. No alternative leads on any target.
The other schedulers were not re-run under them.}
\label{tab:reward}
\setlength{\tabcolsep}{3pt}
\begin{tabular}{@{}lrrr@{}}
\toprule
target & {\footnotesize \texttt{coverage\_interval}} & {\footnotesize \texttt{frontier\_branch}} & {\footnotesize \texttt{rarity\_density}} \\
\midrule
libpng & \bestcell{4.00} & 3.40 & 3.40 \\
libtiff & \bestcell{5.00} & 4.20 & 4.40 \\
openssl & \bestcell{2.00} & 1.40 & 1.60 \\
poppler & \bestcell{4.20} & 3.60 & 3.60 \\
sqlite3 & \bestcell{4.20} & 3.40 & 4.00 \\
\bottomrule
\end{tabular}
\end{table*}

\subsection{Allocation by scheduler}
\label{sec:appendix:alloc}

\myparagraph{Concentration over time.}
The baseline trace diffuses: the largest fuzzer's share falls from $81\%$ in the
first quarter to $25\%$ in the last on libtiff and from $75\%$ to $39\%$ on
sqlite3. Method~II concentrates as evidence accumulates, moving from $92\%$ to
$97\%$ on libtiff and from $38\%$ to $76\%$ on poppler, while remaining stable on
sqlite3 and libxml2. Method~I's concentration follows its first rotting verdict
(\S\ref{sec:rr:phase}). Thus, each scheduler learns an allocation, but its evidence
source differs: reward summaries for the baseline, an expertise-guided phase for
Method~I, and the learned context-to-reward relationship for Method~II.

\begin{figure*}[t]
\centering
\pgfplotsset{
  alloc/.style={
    width=0.27\textwidth, height=2.9cm,
    xmin=0.7, xmax=4.3, xtick={1,2,3,4}, xticklabels={Q1,Q2,Q3,Q4},
    ymin=15, ymax=105, ytick={25,50,75,100},
    xlabel={campaign quarter}, xlabel style={font=\normalsize, yshift=2pt},
    axis lines*=left, tick align=outside,
    ymajorgrids, grid style={draw=black!12},
    xticklabel style={font=\normalsize}, yticklabel style={font=\normalsize},
  },
  base/.style ={cBase, dashed, line width=0.9pt},
  m1/.style   ={cM1, dash pattern=on 3pt off 1.5pt, line width=1.0pt},
  m2/.style   ={cM2, line width=1.2pt},
}
\begin{tikzpicture}
\begin{axis}[alloc, ylabel={largest fuzzer's share (\%)},
             ylabel style={font=\normalsize}, title={\normalsize libtiff}]
\addplot[base] coordinates {(1,80.6) (2,40.8) (3,32.1) (4,24.5)};
\addplot[m1]   coordinates {(1,81.4) (2,68.0) (3,38.1) (4,50.2)};
\addplot[m2]   coordinates {(1,92.1) (2,92.1) (3,95.8) (4,96.6)};
\end{axis}
\end{tikzpicture}\hfill
\begin{tikzpicture}
\begin{axis}[alloc, title={\normalsize poppler}]
\addplot[base] coordinates {(1,29.3) (2,33.7) (3,36.9) (4,32.7)};
\addplot[m1]   coordinates {(1,57.3) (2,61.9) (3,53.0) (4,46.3)};
\addplot[m2]   coordinates {(1,37.6) (2,67.7) (3,71.5) (4,76.4)};
\end{axis}
\end{tikzpicture}\hfill
\begin{tikzpicture}
\begin{axis}[alloc, title={\normalsize sqlite3}]
\addplot[base] coordinates {(1,74.7) (2,62.9) (3,40.6) (4,38.6)};
\addplot[m1]   coordinates {(1,90.0) (2,90.7) (3,79.5) (4,75.9)};
\addplot[m2]   coordinates {(1,66.0) (2,76.5) (3,64.2) (4,63.8)};
\end{axis}
\end{tikzpicture}\hfill
\begin{tikzpicture}
\begin{axis}[alloc, title={\normalsize libxml2}]
\addplot[base] coordinates {(1,59.0) (2,65.6) (3,58.4) (4,49.5)};
\addplot[m1]   coordinates {(1,81.1) (2,93.2) (3,83.5) (4,77.1)};
\addplot[m2]   coordinates {(1,58.2) (2,70.5) (3,60.7) (4,57.1)};
\end{axis}
\end{tikzpicture}
\\[-1pt]
\begin{tikzpicture}[font=\normalsize]
\draw[cBase, dashed, line width=0.9pt] (0,0) -- (0.55,0);
\node[anchor=west] at (0.62,0) {\base};
\draw[cM1, dash pattern=on 3pt off 1.5pt, line width=1.0pt] (3.4,0) -- (3.95,0);
\node[anchor=west] at (4.02,0) {\rrshort};
\draw[cM2, line width=1.2pt] (7.0,0) -- (7.55,0);
\node[anchor=west] at (7.62,0) {\ctxshort};
\end{tikzpicture}
\caption{Allocation concentration by campaign quarter. Each point is the share of
scheduling decisions received by the most-selected fuzzer in that quarter, averaged over
the \nTrials campaigns in the corresponding cell of Table~\ref{tab:main}; the fuzzer at
the peak may change between quarters. Method~I's allocation follows its first rotting
verdict, while Method~II's follows the context-to-reward relationship learned during the
campaign (\S\ref{sec:rr:phase}).}
\label{fig:alloc}
\end{figure*}
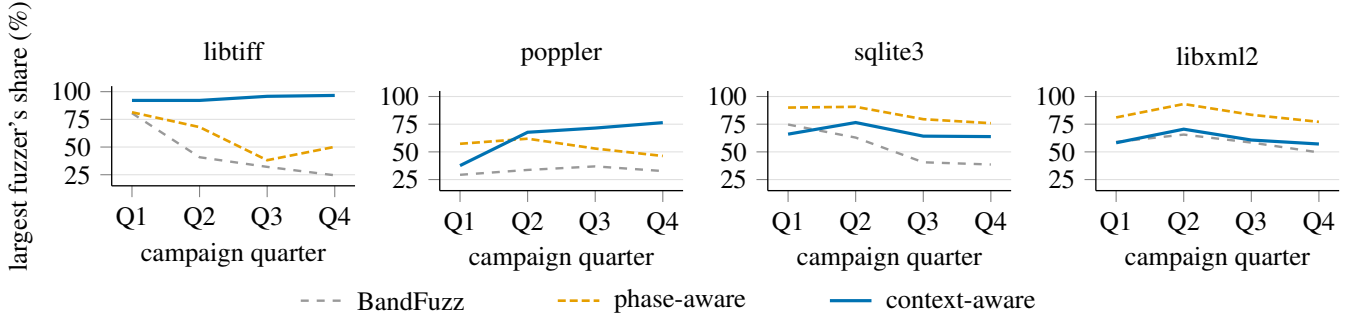

\subsection{Ablation detail}
\label{sec:appendix:ablation}

\myparagraph{Measurement scope.}
The leave-one-feature-out, transfer, pull-count, and low-rank tests use
leave-one-campaign-out folds on all \nTargets targets. The feature-influence replay
in Table~\ref{tab:decision-impact} reports six targets, while the decision totals in
\S\ref{sec:ctx:features} cover all \nTargets. The reward sweep covers five targets,
the linear comparison covers two, and the full-campaign context ablation covers all
\nTargets at \nTrials campaigns per cell.

\myparagraph{Full context against reward history.}
For each fuzzer, we fit a model on all but one campaign and score the held-out
campaign, once with the full context and once with the reward-history signals
\texttt{win\_mean}, \texttt{win\_var}, \texttt{slope}, \texttt{mk\_z}, the three
horizon ratios, and \texttt{cov\_velocity}. The full context has lower error on all
\nTargets targets after Holm correction and higher error on none. It wins on
$66.7\%$ to $96.2\%$ of each target's held-out campaigns; reward history alone has
$2.1\%$ more squared error at the median fold.

Because \texttt{ctx\_unc} is one of the removed signals, the intercept-only
scheduler also makes the context-derived confidence width identical across fuzzers.
Its full-campaign difference therefore measures the complete context-aware decision
signal, including that uncertainty term. Held-out prediction with an intercept and
\texttt{ctx\_unc} alone separates the remaining contextual information.

\myparagraph{Context beyond pull count.}
The \texttt{ctx\_unc} signal is a function of a fuzzer's pull count, an observable
already maintained without a context. An intercept-plus-\texttt{ctx\_unc} model has
$9.7\%$ more held-out prediction error at the median fold than the full context. It
performs worse on all \nTargets targets after Holm correction and better on none.
The predictive contribution of the context therefore extends beyond its
pull-count-derived uncertainty signal.

\myparagraph{Context representation.}
We evaluate reward-supervised low-rank projections~\cite{cai2023doubly}. For each
leave-one-campaign-out fold, we fit the per-fuzzer weight matrix
$\Theta\in\mathbb{R}^{\narms\times\nContextFeatures}$, retain its top-$r$ right
singular directions, refit the ridge model on the projected columns, and compare
held-out squared error with the unprojected model. One target and one rank form a
two-sided Wilcoxon signed-rank test, Holm-corrected over six ranks within the target.

The $532$ campaigns yield $54$ tests and $3{,}189$ paired observations; three folds
lack enough rows for one projected fit. No tested rank improves on the full context
on any target. Thirty-four tests significantly favor the full context, spanning
eight targets, and $72.0\%$ of paired observations favor it. The direction remains
when the projection selects its ridge from $\{1,3,10\}$. In the live comparison, a
rank-3 scheduler reduces mean unique bugs by $0.20$ per target, within the pair's
resolution. Method~II therefore retains the complete context.

\myparagraph{Cost.}
One decision standardizes \nEngines contexts, expands each representation, scores
the fuzzers, and draws a selection in $8.0$\,\textmu s. This cost is negligible
relative to the $\budget=120$\,s turn it governs.

\subsection{Linear against non-linear}
\label{sec:ablation:kernel}

We compare the two forms at the data volume available within a live campaign. Each is
fitted on one fuzzer's first $n$ updates and scored on its next 25. The random-Fourier
form has lower held-out error at every volume tested: on sqlite3 by $36.1\%$, $25.2\%$,
$19.2\%$, $9.1\%$, and $6.9\%$ at $n=12$, $25$, $40$, $100$, and $240$, and on
poppler by $34.2\%$, $30.1\%$, and $32.0\%$ at the first three volumes. Both forms use
the same input, standardization, and $\pm 5$ clamp and have nearly the same number of
columns (16 and 17). They differ in column scale: a linear column is bounded by the
clamp, whereas a random-Fourier column is bounded by $\sqrt{2/\rff}=0.354$. Under the
same ridge penalty, a coefficient therefore moves a linear prediction three to eleven
times farther than a random-Fourier prediction; the corresponding feature-vector norms
are about $4$--$19.4$ and about $1$. These measurements support the bounded
random-Fourier representation used by Method~II, not a claim about nonlinear
expressiveness. The live comparison is $-0.20$ unique bugs per target, within the pair's
resolution (Figure~\ref{fig:ablation}). A scale-matched linear representation or a
linear model with its own ridge penalty was not measured.

\subsection{How the context was chosen}
\label{sec:appendix:featsel}

\myparagraph{Step 1: enumerate and prune.} Two rules reduce the 21 candidates of
\S\ref{sec:ctx:features} to \nContextFeatures. Across $1{,}662{,}902$ decision rows from
the 540 campaigns with context vectors, Rule~1 removes the one candidate with exactly
zero within-campaign range, the covered fraction of the map. Rule~2 removes five exact
arithmetic identities of retained quantities, each with zero residual in double
precision. Table~\ref{tab:features} lists the resulting representation. We then evaluate
each retained feature with leave-one-campaign-out prediction. For every target and
feature, we fit the per-fuzzer ridge on the target's other campaigns with all
\nContextFeatures\ features and with that feature removed, then compare reward-prediction
error on the held-out campaign. A two-sided Wilcoxon signed-rank test uses one paired
observation per campaign, with Holm correction over features within each target.

\myparagraph{What the test says.} The evaluation covers $135$ target--feature tests over
$532$ campaigns. Removing a feature makes prediction significantly worse in 28 tests and
significantly better in 6. No feature is better removed on more than two targets, and the
features that help on some targets overlap those that hurt on others. We therefore retain
the fixed representation instead of selecting it on the campaigns reported here.
\texttt{horizon8\_ratio} is better removed on two targets and better retained on none.
The security signals \texttt{g\_sec} and \texttt{g\_bug} do not improve held-out
prediction on any target, and removing each helps on one. Their influence on flat targets
can still break weakly determined ties, as Step~2 shows.

\myparagraph{Step 2: rank by feature influence.} We replay every logged decision twice,
once from the recorded contexts of all $\narms$ fuzzers and once with one feature replaced
by its standardized mean of zero, while holding the draw fixed. A feature's
\emph{influence} is the fraction of decisions whose top-ranked fuzzer changes. The
ranking reproduces across independent campaign subsets in the style of stability
selection~\cite{meinshausen2010stability}; it is a diagnostic rather than a filter.

\begin{table}[t]
\centering
\tabfont
\caption{How far each of the four strongest context features governs the decision, named
as in Table~\ref{tab:features}: the percentage of logged decisions whose selected fuzzer
changes when that one feature is replaced by its mean. Bold marks each target's own
strongest signal. Neither its identity nor how far it governs the decision is constant
across targets, which is why the weighting is learned per target.}
\label{tab:decision-impact}
\setlength{\tabcolsep}{3pt}
\begin{tabular}{@{}lrrrrrr@{}}
\toprule
feature & \rotninety{libxml2} & \rotninety{libtiff} & \rotninety{libpng}
        & \rotninety{php} & \rotninety{libsndfile} & \rotninety{lua} \\
\midrule
\texttt{g\_rarity} & \bestcell{20.7} & \bestcell{62.4} & 51.5 & \bestcell{59.3} & 20.9 & \bestcell{36.4} \\
\texttt{mk\_z} & 17.1 & 47.8 & 46.9 & 55.4 & 14.5 & 7.9 \\
\texttt{rounds\_since\_improve} & 6.9  & 20.1 & \bestcell{56.3} & 42.2 & 25.3 & 3.4 \\
\texttt{g\_sec} & 3.7  & 8.6  & 14.3 & 37.0 & \bestcell{44.8} & 10.6 \\
\bottomrule
\end{tabular}
\end{table}

Table~\ref{tab:decision-impact} gives the four strongest features. The influence of the
strongest ranges from $20.7\%$ of decisions on libxml2 to $62.4\%$ on libtiff.
\texttt{g\_sec} changes many rankings on libsndfile despite carrying no held-out
predictive signal. libsndfile and php are the two targets on which no scheduler's bug
count differs (\S\ref{sec:eval:rq1}); weakly determined coefficients on these targets
allow a feature to break ties without improving prediction.

\myparagraph{Step 3: is one weighting enough?} We fit one per-fuzzer model on the other
eight targets, freeze it, and compare it on this target's held-out campaign with the
model fitted within this target. Across the same $532$ campaigns, the transferred
weighting predicts significantly worse on eight of \nTargets targets after Holm
correction and better on none. On libtiff it is worse at the median without reaching
significance. It is worse on $66.7\%$ to $97.4\%$ of each target's held-out campaigns.
Replaying the held-out campaigns converts the prediction comparison into decisions: the
transferred and within-target weightings select different top-ranked fuzzers on
$26.8\%$ to $60.4\%$ of a target's decisions, median $51.0\%$. A fixed weighting
therefore fails to transfer, and the difference reaches the scheduling decision.

\subsection{What the difference is made of}
\label{sec:appendix:rq3}

\myparagraph{How far the assumed shape carries.} \S\ref{sec:eval:rq3} names two
measurements of target heterogeneity. First, non-zero rewards occupy $0.4\%$ of libpng's
second-half turns and $40.3\%$ of poppler's. Second, we fit each campaign's per-fuzzer
coefficient matrix on its first half and score it on its second. We compare that error
with the distribution obtained by fitting and scoring random halves of the same
campaign; the ordered split fails when its error exceeds the 95th percentile. The fit
fails in $49.0\%$ of sqlite3 campaigns, $37.0\%$ of libxml2 campaigns, and $35.9\%$ of
poppler campaigns. The rates on the other six targets are $9.5\%$ for libtiff, $7.9\%$
for lua, $5.9\%$ for openssl, $2.5\%$ for libsndfile, $0.7\%$ for libpng, and zero for
php. Thus both the remaining reward and within-campaign drift vary sharply by target.
Their rank correlations with Method~II's gain are $0.20$ and $0.00$
(\S\ref{sec:appendix:motivcaveat}); the direct Method~I--Method~II comparison determines
which design performs better.

\myparagraph{Plateau signature.} We call a signal moved when its standardized value
changes by more than one standard deviation across a window. For each trigger of a bug
that \ctx reaches in more campaigns than \base, we take one window per fuzzer over the
preceding 90 minutes. \texttt{rounds\_since\_improve} moves in $65\%$ of windows (mean
standardized change $+0.61$), \texttt{time\_since\_run} in $53\%$ ($+0.39$), and
\texttt{cov\_velocity} in $12\%$ while remaining below its campaign average. Over all
bugs, the coverage signal moves more often and the plateau signals less often. These
measurements describe trigger windows but do not supply a time-matched base rate. In
particular, \texttt{time\_since\_run} rises with idleness. They therefore establish a
contrast between the observed sets, not unusual plateau activity.
\texttt{rounds\_since\_improve} is nevertheless observable context supplied to the
model (Table~\ref{tab:features}).

\subsection{Every fuzzer alone}
\label{sec:appendix:single}

\myparagraph{Single-fuzzer protocol.} Each single-fuzzer campaign uses the same dispatch
loop, container image, target build, evaluator, shared seed store, Magma seed corpus,
memory limit, \nEngines worker cores, \campaignHours-hour duration, and $120$-second turns
as the scheduler campaigns. Its roster contains one fuzzer. We run \nTrials campaigns
for each of the $54$ target--fuzzer pairs.

\subsection{Sampling and association}
\label{sec:appendix:motivcaveat}

A campaign enters the ordering measurement only when four fuzzers complete four turns in
both windows of a pair. This selection favors less concentrated schedules. The same
filter admits $1{,}841$ campaign--fuzzer pairs to the decay measurement. Across the
\nTargets targets, the rank correlation between reward decay and Method~II's gain is
$0.20$; the correlation between within-window ranking agreement and that gain is $0.00$.
Two of the nine targets are flat, with no scheduler difference. These measurements do
not establish a dose response between either motivating statistic and scheduler gain.

\subsection{The real-world harnesses}
\label{sec:appendix:realworld}

\myparagraph{Observed failure modes.} Across all \nRealDefects defects, $56$ fault at an address
below \texttt{0x10000}, a page that cannot be mapped. $8$ perform a write access of any
kind and $4$ of those write past an allocation. The rest fault without writing anything
we observed. These run-level observations differ from the weakness classes in
Table~\ref{tab:weakness}, so the write count is lower than the table's
out-of-bounds-write row. They also do not partition the two disclosure channels of
\S\ref{sec:eval:realworld}, whose criterion is the worst outcome observed for a
defect across its runs rather than the address of one fault, and therefore do not add
to \nRealPublic and \nRealPrivate.

\myparagraph{Targets and harnesses.} We evaluate \nRealPrograms widely used open-source
C++ programs that parse untrusted files through documented entry points. Each harness is
a minimal driver for one such entry point and defines only its own \texttt{main}:
\begin{itemize}\setlength{\itemsep}{0pt}\setlength{\parskip}{0pt}\raggedright
\item MNN: the converter's format front ends, each reached by
  \texttt{MNN::Cli::convertModel} on a model of that format;
\item Arm NN: \texttt{IDeserializer::\allowbreak CreateNetworkFromBinary};
\item ncnn: \texttt{ncnn::Net::load\_param};
\item FreeCAD: \texttt{MeshInput::LoadFormat};
\item Krita: \texttt{loadFromDevice}, \texttt{KisAslReader::\allowbreak readFile}
  and \texttt{Compression::\allowbreak uncompress}.
\end{itemize}
Each target uses the revision in Table~\ref{tab:realworld}.

\myparagraph{From crashes to defects.} We minimize crashing inputs and deduplicate them
on faulting function, file, line, fault type, and entry point. Three sites in
Tables~\ref{tab:locations} and~\ref{tab:locationsb} contain multiple records: two are
separated by fault type and one by entry point. We remove candidates that fail to
reproduce without sanitizers and coverage instrumentation or already appear in the
project's tracker, change history, or CVE record. \nRealDefects defects remain. Each is
symbolized to a function, file, and line and reconfirmed in five of five runs on a build
without coverage instrumentation or sanitizers.

\subsection{Immediate-reward objective}
\label{sec:appendix:setting}

A campaign-optimal policy would model the shared seed store and the transition value of
each decision. The dispatch loop observes per-fuzzer coverage volume, a lossy projection
under which campaigns with different stores can look identical. Method~II therefore
selects the fuzzer with the highest predicted next-turn reward. The retained-rate
measurement of \S\ref{sec:motivation:decay} and the alternative rewards in
Table~\ref{tab:reward} test the direction of this objective. No alternative reverses the
ordering on the tested benchmark. The evidence supports the measured immediate-reward
direction, not campaign-level optimality.

\subsection{Full-history weighting}
\label{sec:appendix:gamma}

The exponential weighting assigns an observation $k$ decisions old weight $\gamma^k$.
The swept settings $\gamma=0.995$ and $0.99$ have half-lives of approximately $139$ and
$69$ decisions. Both produce significantly worse held-out prediction on all \nTargets
targets after Holm correction and better prediction on none. Method~II therefore retains
the full campaign history with $\gamma=1$.

\subsection{Selection-rule scale and assumptions}
\label{sec:appendix:select}

At the deployed constants, the standard self-normalized calculation uses
$\rff+1=17$ parameters per fuzzer, $\ridge=10$, $\|\featmap\|^2\leq 3$ including the
intercept, reward in $[0,1]$, $\tfrac12$-sub-Gaussian noise, $\|\theta\|\leq 1$, and
$\delta=0.05$. It gives a regret bound near $12{,}200$ over the approximately $4{,}320$
decisions in a \campaignHours-hour campaign, above the trivial $4{,}320$ reward bound.
It first falls below that trivial bound at approximately $76{,}000$ decisions, more than
17 campaigns. The bound is therefore uninformative at the evaluated horizon.

The deployed setting also has \nEngines concurrent selections whose rewards arrive after
the $\budget$-second turn, and the context--reward map can drift within a campaign
(\S\ref{sec:appendix:rq3}). We evaluate the selection rule empirically under this
concurrency, delay, and drift. Its width is
$u_\arm=\sqrt{\featmap_\arm^{\!\top}A_\arm^{-1}\featmap_\arm}$. The basis width $\rff$,
bandwidth $\sigma$, and ridge $\ridge$ affect this quantity through the effective
dimension of $A_\arm$ and the rate at which the kernel decorrelates contexts; only
$\ridge$ is swept. The ridge also makes
$A_\arm=\ridge I+\sum\featmap\featmap^{\!\top}$ invertible from the first observation
and shrinks a fuzzer's $\rff+1$ parameters while data is scarce.

\subsection{The main table in full}
\label{sec:appendix:mainfull}

\begin{table*}[t]
\centering
\scriptsize
\caption{Each of this paper's schedulers against each baseline, per target.
$\hat{A}_{12}$ is the Vargha--Delaney effect size, the probability that a randomly chosen
campaign of ours triggers more bugs than a randomly chosen campaign of that scheduler,
ties counted as half. $p_{\mathrm{H}}$ is a two-sided Mann--Whitney U test after Holm
correction over the family of \nTargets targets, one family per pair.
\textbf{Bold} marks $p_{\mathrm{H}} < 0.05$.}
\label{tab:main-full}
\setlength{\tabcolsep}{4pt}
\begin{tabular}{@{}l rr rr rr rr rr rr@{}}
\toprule
& \multicolumn{6}{c}{\ctx} & \multicolumn{6}{c}{\rr} \\
\cmidrule(lr){2-7}\cmidrule(lr){8-13}
& \multicolumn{2}{c}{vs \base} & \multicolumn{2}{c}{vs \autofz}
& \multicolumn{2}{c}{vs \legion}
& \multicolumn{2}{c}{vs \base} & \multicolumn{2}{c}{vs \autofz}
& \multicolumn{2}{c}{vs \legion} \\
\cmidrule(lr){2-3}\cmidrule(lr){4-5}\cmidrule(lr){6-7}
\cmidrule(lr){8-9}\cmidrule(lr){10-11}\cmidrule(lr){12-13}
target & $\hat{A}_{12}$ & $p_{\mathrm{H}}$ & $\hat{A}_{12}$ & $p_{\mathrm{H}}$ & $\hat{A}_{12}$ & $p_{\mathrm{H}}$ & $\hat{A}_{12}$ & $p_{\mathrm{H}}$ & $\hat{A}_{12}$ & $p_{\mathrm{H}}$ & $\hat{A}_{12}$ & $p_{\mathrm{H}}$ \\
\midrule
libpng & 0.80 & \textbf{0.025} & 0.90 & \textbf{0.0031} & 0.70 & 0.202 & 0.75 & 0.226 & 0.85 & \textbf{0.023} & 0.65 & 1.000 \\
libsndfile & 0.50 & 1.000 & 0.50 & 1.000 & 0.50 & 1.000 & 0.50 & 1.000 & 0.50 & 1.000 & 0.50 & 1.000 \\
libtiff & 1.00 & \textbf{0.0004} & 1.00 & \textbf{0.0001} & 1.00 & \textbf{0.0004} & 0.70 & 0.595 & 0.60 & 1.000 & 0.74 & 0.385 \\
libxml2 & 0.96 & \textbf{0.0012} & 0.74 & 0.264 & 0.92 & \textbf{0.0035} & 0.67 & 0.775 & 0.52 & 1.000 & 0.63 & 1.000 \\
lua & 0.75 & 0.101 & 0.70 & 0.331 & 0.60 & 1.000 & 0.56 & 1.000 & 0.51 & 1.000 & 0.42 & 1.000 \\
openssl & 0.70 & 0.101 & 0.50 & 1.000 & 0.70 & 0.202 & 0.60 & 1.000 & 0.40 & 1.000 & 0.60 & 1.000 \\
php & 0.50 & 1.000 & 0.50 & 1.000 & 0.50 & 1.000 & 0.50 & 1.000 & 0.50 & 1.000 & 0.50 & 1.000 \\
poppler & 0.88 & \textbf{0.017} & 0.89 & \textbf{0.012} & 0.71 & 0.382 & 0.55 & 1.000 & 0.65 & 1.000 & 0.39 & 1.000 \\
sqlite3 & 0.84 & \textbf{0.022} & 1.00 & \textbf{0.0003} & 1.00 & \textbf{0.0003} & 0.45 & 1.000 & 0.65 & 0.614 & 0.65 & 0.614 \\
\midrule
summed mean & \multicolumn{2}{c}{$+20.4\%$} & \multicolumn{2}{c}{$+18.8\%$} & \multicolumn{2}{c}{$+17.9\%$} & \multicolumn{2}{c}{$+7.4\%$} & \multicolumn{2}{c}{$+5.9\%$} & \multicolumn{2}{c}{$+5.2\%$} \\
\bottomrule
\end{tabular}
\end{table*}

\begin{table*}[t]
\centering
\tabfont
\caption{Every fuzzer alone, mean unique Magma bugs over \nTrials \campaignHours-hour campaigns per cell, on the same harness, seeds, cores and budget as every scheduler in Table~\ref{tab:main}. The last row is the strongest of them on each target, a choice made after the campaigns have run. \textbf{Bold} marks the highest mean in a column.}
\label{tab:single}
\setlength{\tabcolsep}{4pt}
\begin{tabular}{@{}l rrrrrrrrr r@{}}
\toprule
fuzzer & libpng & libsndfile & libtiff & libxml2 & lua & openssl & php & poppler & sqlite3 & sum \\
\midrule
\aflpp & 2.20 & 7.00 & \bestcell{4.40} & \bestcell{2.80} & \bestcell{1.40} & \bestcell{2.00} & 2.40 & 3.00 & \bestcell{3.20} & 28.40 \\
\darwin & 1.60 & 7.00 & 3.00 & 2.40 & 1.00 & 1.60 & 3.00 & 3.00 & 1.20 & 23.80 \\
\honggfuzz & \bestcell{4.00} & 1.20 & 3.00 & 2.00 & 1.00 & 1.60 & 2.80 & 2.20 & 2.00 & 19.80 \\
\lafintel & 2.20 & 7.00 & 4.00 & 2.60 & 1.20 & \bestcell{2.00} & 3.00 & 2.80 & 3.00 & 27.80 \\
\mopt & 2.60 & 7.00 & 3.20 & 2.00 & 1.00 & \bestcell{2.00} & 3.00 & 3.00 & 2.80 & 26.60 \\
\radamsa & 1.40 & 6.40 & 1.60 & 1.20 & \bestcell{1.40} & 1.00 & 3.00 & \bestcell{4.20} & 1.60 & 21.80 \\
\midrule
\emph{strongest, per target} & 4.00 & 7.00 & 4.40 & 2.80 & 1.40 & 2.00 & 3.00 & 4.20 & 3.20 & \emph{32.00} \\
\bottomrule
\end{tabular}
\end{table*}

\begin{table*}[t]
\centering
\tabfont
\caption{Standard deviation of unique bugs across the \nTrials campaigns of each cell of Table~\ref{tab:main}. \textbf{Bold} marks the largest spread in a column.}
\label{tab:main-sd}
\setlength{\tabcolsep}{4pt}
\begin{tabular}{@{}l rrrrrrrrr@{}}
\toprule
scheduler or fuzzer & libpng & libsndfile & libtiff & libxml2 & lua & openssl & php & poppler & sqlite3 \\
\midrule
\ctxshort & 0.00 & 0.00 & 0.00 & 0.42 & 0.52 & 0.00 & 0.00 & 0.79 & 0.42 \\
\rrshort & 0.32 & 0.00 & \bestcell{0.82} & \bestcell{0.95} & \bestcell{0.67} & 0.42 & 0.00 & 0.63 & 0.48 \\
\midrule
\base & 0.52 & 0.00 & 0.48 & 0.32 & 0.32 & 0.52 & 0.00 & 0.57 & 0.52 \\
\autofz & 0.42 & 0.00 & 0.00 & 0.84 & 0.42 & 0.00 & 0.00 & 0.79 & 0.00 \\
\legion & 0.52 & 0.00 & 0.52 & 0.42 & 0.52 & 0.52 & 0.00 & \bestcell{0.84} & 0.00 \\
\midrule
\aflpp & 0.42 & 0.00 & 0.52 & 0.79 & 0.52 & 0.00 & \bestcell{1.26} & 0.00 & 0.42 \\
\darwin & 0.52 & 0.00 & 0.00 & 0.52 & 0.00 & 0.52 & 0.00 & 0.00 & \bestcell{1.23} \\
\honggfuzz & 0.00 & 0.42 & 0.00 & 0.00 & 0.00 & 0.52 & 0.42 & 0.42 & 0.00 \\
\lafintel & 0.42 & 0.00 & 0.67 & 0.52 & 0.42 & 0.00 & 0.00 & 0.42 & 0.00 \\
\mopt & 0.52 & 0.00 & 0.42 & 0.00 & 0.00 & 0.00 & 0.00 & 0.00 & 0.79 \\
\radamsa & 0.52 & \bestcell{0.52} & 0.52 & 0.42 & 0.52 & 0.00 & 0.00 & 0.79 & 1.07 \\
\bottomrule
\end{tabular}
\end{table*}

\begin{table*}[t]
\centering
\tabfont
\caption{Mean edge coverage for every cell of Table~\ref{tab:main}, reported for reference and never used to decide a comparison. \textbf{Bold} marks the highest mean in a column.}
\label{tab:main-edges}
\setlength{\tabcolsep}{4pt}
\begin{tabular}{@{}l rrrrrrrrr@{}}
\toprule
scheduler or fuzzer & libpng & libsndfile & libtiff & libxml2 & lua & openssl & php & poppler & sqlite3 \\
\midrule
\ctxshort & 1,469 & 2,496 & \bestcell{3,889} & 8,954 & 5,126 & \bestcell{8,500} & \bestcell{3,945} & 22,242 & 13,101 \\
\rrshort & 1,467 & \bestcell{2,578} & 3,834 & 8,452 & \bestcell{5,159} & 8,450 & 3,944 & 22,189 & 11,903 \\
\midrule
\base & 1,465 & 2,509 & 3,729 & 8,810 & 5,118 & 8,468 & 3,945 & 22,132 & 12,272 \\
\autofz & 1,467 & 2,462 & 3,714 & \bestcell{9,210} & 5,127 & 8,469 & 3,944 & 22,089 & 11,597 \\
\legion & 1,463 & 2,439 & 3,669 & 8,276 & 5,144 & 8,452 & 3,944 & 22,201 & 11,236 \\
\midrule
\aflpp & 1,194 & 2,527 & 3,881 & 8,544 & 5,103 & 8,477 & 3,740 & 22,180 & \bestcell{13,269} \\
\darwin & 1,113 & 2,515 & 3,207 & 6,396 & 5,067 & 8,452 & 3,907 & 22,344 & 6,844 \\
\honggfuzz & \bestcell{1,469} & 1,675 & 3,081 & 5,794 & 4,911 & 8,419 & 3,942 & 22,023 & 9,581 \\
\lafintel & 1,203 & 2,481 & 3,811 & 8,054 & 5,094 & 8,447 & 3,928 & 22,032 & 12,111 \\
\mopt & 1,222 & 2,565 & 3,719 & 7,932 & 5,103 & 8,484 & 3,930 & \bestcell{22,547} & 11,282 \\
\radamsa & 1,026 & 2,365 & 2,848 & 7,007 & 5,120 & 8,440 & 3,904 & 21,977 & 8,260 \\
\bottomrule
\end{tabular}
\end{table*}

\begin{table*}[t]
\centering
\tabfont
\caption{Mean injected bugs \emph{reached} per campaign for every cell of Table~\ref{tab:main}, from the Magma oracle's own reached record. A bug is reached when a campaign executed its site, whether or not the input that got there met the condition that trips it, so this column bounds the bug counts of Table~\ref{tab:main} from above and Table~\ref{tab:ceilings} bounds it in turn. \textbf{Bold} marks the highest mean in a column.}
\label{tab:main-reached}
\setlength{\tabcolsep}{4pt}
\begin{tabular}{@{}l rrrrrrrrr@{}}
\toprule
scheduler or fuzzer & libpng & libsndfile & libtiff & libxml2 & lua & openssl & php & poppler & sqlite3 \\
\midrule
\ctxshort & 6.00 & 8.00 & \bestcell{7.90} & 8.50 & 4.00 & 4.00 & 5.00 & 16.00 & \bestcell{12.10} \\
\rrshort & 6.00 & 8.00 & 7.40 & 8.40 & 4.00 & 4.00 & 5.00 & 15.90 & 11.20 \\
\midrule
\base & 6.00 & 8.00 & 6.90 & 8.10 & 4.00 & 4.00 & 5.00 & 15.60 & 11.40 \\
\autofz & 6.00 & 8.00 & 6.80 & 8.20 & 4.00 & 4.00 & 5.00 & 15.60 & 10.60 \\
\legion & 6.00 & 8.00 & 6.80 & 8.20 & 4.00 & 4.00 & 5.00 & 16.00 & 10.20 \\
\midrule
\aflpp & 6.00 & 8.00 & 7.20 & \bestcell{8.60} & 4.00 & 4.00 & 4.00 & 16.00 & 11.20 \\
\darwin & 6.00 & 8.00 & 5.40 & 5.40 & 4.00 & 4.00 & 5.00 & 16.00 & 8.60 \\
\honggfuzz & 6.00 & 7.00 & 5.00 & 5.00 & 3.00 & 4.00 & 5.00 & 15.00 & 10.00 \\
\lafintel & 6.00 & 8.00 & 7.00 & 8.40 & 4.00 & 4.00 & 5.00 & 16.00 & 10.80 \\
\mopt & 6.00 & 8.00 & 6.40 & 7.60 & 4.00 & 4.00 & 5.00 & 16.00 & 10.20 \\
\radamsa & 6.00 & 8.00 & 5.00 & 7.60 & 4.00 & 4.00 & 5.00 & 15.80 & 8.80 \\
\bottomrule
\end{tabular}
\end{table*}

\subsection{The phase-aware scheduler in detail}
\label{sec:appendix:rr}

\myparagraph{The horizon ladder, rung by rung.} With $\explore = 1$ and a reward in
$[0,1]$, $h = 2$ is selected over $h$ unless the last $h$ rewards average more than
$\sqrt{\log t}\,(1/\sqrt{2} - 1/\sqrt{h})$ above the last two. Since two means of a
$[0,1]$ reward differ by at most $1$, horizon $h$ becomes unreachable once
$\log t > (1/\sqrt{2} - 1/\sqrt{h})^{-2}$. Over the $4{,}320$ turns an
\campaignHours-hour campaign holds, that retires $h = \window$ after turn $17$ and
$h = 16$ after turn $120$, and retires neither $h = 8$, which would need $t > 2{,}981$,
nor $h = 4$, which would need $t > 1.3 \times 10^{10}$. A three-rung
ladder survives. $h = 8$ is reachable only where the last eight rewards average $0.95$
above the last two and $h = 4$ only where the last four average $0.56$ above them. So this
branch reads the mean of the fuzzer's two most recent rewards, except on
decisions where its recent rewards differ that sharply. We have not measured
how often that is. The draw's scale $\sqrt{1/12 + 1/12} = 0.408$ is itself a bound only once a
fuzzer has four rewards to its name, since the largest sample variance of $m$ points in
$[0,1]$ is $m/(4(m-1))$.

Method~I has exactly two phases. Rising sequences use the rising estimator; flat and
downward sequences both use the rotting estimator. The phase is a one-bit context supplied
by domain expertise, and it selects the estimator. Online uncertainty then orders fuzzers
within the selected phase. The two index forms originate in models whose rewards change
with an arm's pull count~\cite{levine2017rotting} or with global
time~\cite{seznec2020single}. Fuzzer reward also changes with inputs published by peers,
so Method~I uses the estimators without importing those models' guarantees.

The gap between a rising and a rotting index is wider than the range of a reward mean.
A draw whose scale is at most $\sqrt{1/12+1/12}=0.408$ crosses that gap only in the
Gaussian tail. Thus, the phase supplies most of the allocation decision and learned
uncertainty orders fuzzers within a phase; the crossing rate was not measured. This
expertise-guided context improves summed mean unique bugs by $5.2\%$--$7.4\%$ over the
three baselines. Method~II learns from the full context and improves on Method~I by
$12.1\%$, leading on seven targets and tying on two. The comparison shows that the
one-bit prior is useful and that learning from the complete context is stronger.

\subsection{Construction and scope}
\label{sec:appendix:prop}

\myparagraph{The construction in full.}
Take two fuzzers and a shared store. Whichever fuzzer runs on a turn publishes an input
into the store, and that input opens ground new to fuzzer~1 with probability
$\tfrac{1}{2}$ and, independently, ground new to fuzzer~2 with probability
$\tfrac{1}{2}$. Let $z^{(\arm)}_\round \in \{0,1\}$ record whether the store gained,
during turn $\round - 1$, an input opening ground new to fuzzer $\arm$, and let fuzzer
$\arm$ return $\tfrac{1}{2} + \tfrac{1}{2}\mathbf{1}[z^{(\arm)}_\round = 1]$. Covering
ground a peer has just opened raises the reward. The coupling is endogenous, since the
publication is produced by the turn the policy chose to run, and monotone as
\S\ref{sec:motivation:decay} requires, since a publication can only raise a fuzzer's
reward and never lower it. Its law is nonetheless the same under every policy, because
some fuzzer runs on every turn and what the publication opens does not depend on which
fuzzer produced it. $z^{(1)}_\round$ and $z^{(2)}_\round$ are independent fair coins
whatever the schedule. That invariance is a requirement rather than a convenience.
Suppose what a publication opened depended on which fuzzer produced it, say a turn of
fuzzer~2 opened ground new to fuzzer~1 and a turn of fuzzer~1 opened nothing. A policy
that knows its own play history would then know the conditional law of $z_\round$, and
after a turn of fuzzer~1 would know $z^{(1)}_\round = 0$ outright, recovering from its own
record the quantity the construction withholds. What a fuzzer's own history provably
cannot reach is therefore exactly the part of the coupling its own choices do not steer,
and that is the part this instance isolates; on the remainder a context-blind policy is
uninformed rather than provably lost. A policy that does not observe them selects at $\round$ from
quantities measurable with respect to rewards realized before $\round$. Those are
functions of $z_1, \dots, z_{\round-1}$ and independent of $z_\round$, so it earns
$\tfrac{1}{2} + \tfrac{1}{2}\cdot\tfrac{1}{2} = \tfrac{3}{4}$ per turn whatever it does,
whatever it reads of its own history, in whatever order, and however long it
collects. A policy that observes $z_\round$ runs a fuzzer whose indicator is $1$ whenever
one exists, which happens with probability $1 - \tfrac{1}{4}$, and earns
$\tfrac{3}{4}\cdot 1 + \tfrac{1}{4}\cdot\tfrac{1}{2} = \tfrac{7}{8}$. The gap is
$\tfrac{1}{8}$ a turn, so the regret is linear in the horizon.

The proposition scopes the information available to the policy, not the amount of history
it retains. Because $z_\round$ is independent of everything realized before round
$\round$, the order of earlier plays also reveals nothing about it. The construction does
not identify sufficient observables or predict deployed performance. The signals in
\S\ref{sec:ctx:features} measure consequences of the coupling, and
\S\ref{sec:ablation:context} evaluates their contribution. The deployed setting remains
empirical because its selections are concurrent and rewards are delayed
(\S\ref{sec:ctx:assumptions}).

\subsection{The constants outside the sweep}
\label{sec:appendix:constants}

\myparagraph{Fixed constants.} Method~II fixes six constants outside the sweep: basis
width $\rff=16$, bandwidth $\sigma=4$, one warm-up turn per fuzzer, the $\pm5$ clamp,
the 84-decision reward window, and perturbation multiplier one. The window also defines
Method~I's trend test and horizon ladder. Method~II's Gaussian perturbation multiplier is
distinct from Method~I's $\explore$ coefficient. The $\budget=120$\,s turn budget is
inherited from \base and held fixed across schedulers. The reported results do not
establish insensitivity to these constants.

\subsection{The real-world census}
\label{sec:appendix:census}

\begin{table}[t]
\centering
\tabfont
\caption{The \nRealDefects real-world defects, by program. \emph{Found} counts distinct
faulting sites under the rule of \S\ref{sec:eval:realworld}. Every one crashes a build
with no sanitizer and no coverage instrumentation, on five runs out of five. The two
right-hand columns give the class we assigned at triage, from the worst outcome we
observed across a defect's runs. \emph{Crash} is a fault at an unmappable address or an
arithmetic fault. \emph{Withheld} is memory corruption or information disclosure.
Every defect went to its maintainers through the channel that project's policy
prescribes. Of the \nRealPublic in the \emph{crash} column, $42$ carry an issue on the
project's own tracker with a working reproducer. Arm NN's $31$ do not, because that
vendor asked that any public disclosure of its defects be coordinated with it, and we
agreed. No defect in the \emph{withheld} column carries a reproducer here.}
\label{tab:realworld}
\setlength{\tabcolsep}{4pt}
\begin{tabular}{@{}llrrr@{}}
\toprule
program & version & found & crash & withheld \\
\midrule
MNN         & \texttt{19af6da} & 62 & 33 & 29 \\
Arm NN      & v26.07           & 42 & 31 & 11 \\
Krita       & 6.0.3            & 14 &  8 &  6 \\
ncnn        & \texttt{c189d88} &  1 &  1 &  0 \\
FreeCAD     & 1.1.3            &  1 &  0 &  1 \\
\midrule
\multicolumn{2}{@{}l}{\emph{total}} & \bestcell{\nRealDefects} & \nRealPublic & \nRealPrivate \\
\bottomrule
\end{tabular}
\end{table}

\subsection{The reward sweep}
\label{sec:appendix:rewardsweep}

\myparagraph{Alternative rewards.} We sweep two alternatives on five targets with
\nTrials campaigns per cell: \texttt{frontier\_branch}, which credits branches at the
edge of the shared map, and \texttt{rarity\_density}, which credits rare covered edges.
The comparison is directional. Neither alternative leads the shipped reward on any of
the five targets. On libtiff and sqlite3, where Method~II's main result survives
correction, the shipped reward leads both alternatives. The measured direction therefore
persists across these reward definitions; the experiment does not rank reward functions.

\subsection{Full-campaign ablations by pair}
\label{sec:appendix:ablationfull}

Every comparison in Figure~\ref{fig:ablation} is a pair of cells. The pair-specific
resolution is the smallest attainable difference in means whose two-sided permutation
tail is at most $0.05$ (\S\ref{sec:eval:setup}). For the full context against the
intercept-only scheduler, the resolutions are $1.00$ on poppler; $0.80$ on libtiff,
libxml2, and sqlite3; and $0.60$ on libpng and lua. No attainable difference resolves on
openssl, libsndfile, or php. The intercept-only scheduler's target-level gaps over
\base are $-0.20$ on libpng, $0.00$ on libsndfile, $-0.10$ on libtiff, $+0.10$ on
libxml2, $+0.30$ on lua, $0.00$ on openssl, $0.00$ on php, $-0.50$ on poppler, and
$0.00$ on sqlite3. None resolves, and the summed mean is $28.00$ against \base's
$28.40$. Method~II reaches $34.20$, a $+6.20$ gain. Because the control removes
\texttt{ctx\_unc} and its contribution to the width, this comparison measures the
complete context-aware decision rather than semantic context alone.

Removing exploration costs $1.00$ mean unique bug per target: the ablated scheduler sums
to $25.20$ over the \nTargets targets, against Method~II's $34.20$. Without the Gaussian
term, a passed-over fuzzer regains the top prediction zero times across all learned
decisions in this configuration, and the scheduler locks onto one fuzzer.

The settings of \S\ref{sec:ctx} that no row of Figure~\ref{fig:ablation} covers
are named in \S\ref{sec:appendix:constants}.

\subsection{Pairwise permutation resolution}
\label{sec:appendix:resolution}

For each two-cell comparison, we pool its $20$ campaign-level bug counts and
enumerate every assignment of ten observations to each cell. The pair's resolution
is the smallest attainable absolute difference in means whose two-sided
permutation tail is at most $0.05$. Against the baseline with the highest observed
mean on each target, Method~II's resolution is $0.6$ bug on libtiff and lua, $0.8$
on libxml2 and sqlite3, and $1.0$ on poppler. On libpng, openssl, libsndfile, and
php, the permutation distribution admits no attainable threshold at $0.05$.

\subsection{Arrival time in full}
\label{sec:appendix:arrival}

\myparagraph{Arrival time.} The 25 bugs triggered by every campaign of every scheduler
have median first-arrival times within the first $11\%$ of the campaign, with a median of
$2\%$ of the budget. The seven differentiating bugs with in-budget medians range from
$5.6$ to $22.1$ hours, with median $13.6$ hours; five arrive after the first quarter and
one in the final quarter. On
libxml2, the two universal bugs arrive by 15 minutes, while the two differentiating bugs
arrive under \ctx at medians of $13.7$ and $22.1$ hours. \base reaches one in a single
campaign and never reaches the other.

\subsection{Baseline implementations}
\label{sec:appendix:comparison}

We reimplement all three baselines inside the common dispatch loop. For each, we map the
paper's algorithm and, when available, its released code to the implementation rule by
rule, following the paper when the two differ. \base uses the algorithm and constants
of~\cite{shi2025bandfuzz}. \autofz~\cite{fu2023autofz} implements the preparation phase,
unique-path allocation, focus phase, and additive-increase multiplicative-decrease
threshold with the paper's $T_{\mathrm{prep}}$, $T_{\mathrm{focus}}$, initial threshold,
and 30-second short interval. Per-fuzzer unique paths use the shared map's record of the
engine that first covered each edge, which implements the stated intersection
subtraction. \legion~\cite{legion2026} uses 600-second rounds, five seed-evaluation
dimensions, spread-based adaptive weights, and bounded-exploitation softmax. Deep edges
come from the target's call graph and rare edges from the shared map's per-edge hit
counts.

\autofz and \legion assign core shares for a round rather than selecting one fuzzer per
worker request. We convert each share to whole core counts that sum to \nEngines and
assign successive decisions cyclically. Each fuzzer therefore receives exactly its
requested core count over every cycle of \nEngines decisions. Both use the wall clock
specified by their papers: \autofz alternates phases and \legion rescores each round.
All implementations share this paper's reward definition, seed sharing, harnesses, and
execution loop, so the comparison isolates scheduling rules rather than reproducing the
complete systems or their published numbers. We do not tune the baselines or expose the
Magma oracle. The artifact includes the three implementations, rule-correspondence
records, and configuration tests.

\subsection{Exposure paths}
\label{sec:appendix:delivery}

Krita's style, brush, and document readers execute when a user opens or imports a file,
including files received by mail, download, or shared drive. FreeCAD's mesh reader is
exposed through its import path. ncnn's parameter loader executes when an application
loads a model at run time. MNN's converter and Arm NN's deserializer account for $104$ of
the \nRealDefects defects. Their exposure is model supply chain: MNN processes a model
during developer-facing conversion, and Arm NN processes a serialized network when an
application loads it.

\subsection{Ordering-measurement interpretation}
\label{sec:appendix:attenuation}

\myparagraph{The leader reading.} A campaign's two halves name different leaders in
$78.5\%$ of the $144$ pairs and the two halves of a single window in $78.5\%$ as well,
$59.6\%$ against $64.1\%$ at quarters and $35.9\%$ against $39.0\%$ at twelfths. Every
paired difference's interval crosses zero. The unmatched reading is $70.4\%$ of the $203$
against the $82.0\%$ a leader drawn at random among the $5.6$ fuzzers this measurement
admits on average would give. It is coarser than the rank correlation because it discards
everything below first place, which is why \S\ref{sec:motivation:decay} reports it for
direction only.

Adjacent-window agreement and within-window agreement are both low, and their paired
differences are statistically indistinguishable. Correcting their ratio for attenuation
gives $0.95$ at halves, $0.82$ at quarters, and $1.0$ at twelfths, but each ratio divides
two small, noisy quantities; the half-window paired difference has a 95\% interval of
$[-0.08,0.09]$. The measurement therefore cannot distinguish a changing true ranking
from a stable ranking estimated poorly at this timescale. The main text reports this
indistinguishability.

Method~II differs from \base in two relevant ways: it pools the full history with
$\gamma=1$ and conditions predictions on context. The intercept-only control also pools
the history but removes both semantic context and the context-derived confidence width
(\S\ref{sec:ablation:context}); it isolates the complete context-aware decision rather
than conditioning alone.

\subsection{Compute}
\label{sec:appendix:compute}

Every benchmark campaign ran for \campaignHours hours on six cores of one 188-core
machine. The real-world campaigns ran on the same machine without a fixed duration. The
real-world evaluation reports defects found, not a discovery rate.

\end{document}